\documentclass[a4paper,fleqn,10pt]{article}
\usepackage{amsmath}
\usepackage{amssymb}
\usepackage{array}
\usepackage{calc}
\usepackage{longtable}
\usepackage{multirow}
\usepackage{axodraw}
\usepackage{pstricks}
\usepackage{graphicx}
\usepackage{xspace}
\usepackage{cite}
\usepackage{mcite}
\usepackage[a4paper,pdfborder={0 0 0}]{hyperref}
\usepackage[format=hang,labelfont=bf,hypcap=true]{caption}
\usepackage{subfig}
\usepackage{sectsty}
\allsectionsfont{\sffamily}
\subsubsectionfont{\mdseries\itshape\large}
\let\spreprint\empty
\newcommand{\preprint}[1]{\def\spreprint{\protect#1}}
\let\sinstitute\empty
\newcommand{\institute}[1]{\def\sinstitute{\protect#1}}
\makeatletter
\renewcommand{\maketitle}{\begingroup
  \null\thispagestyle{empty}%
    \ifx\spreprint\empty
      \vskip 5ex
    \else
      \flushright\large\spreprint\vskip 2ex
    \fi
    \vskip 5ex
    \flushleft
      {\sffamily\bfseries\huge\@title}\vskip 2ex
      \@author\vskip 2ex
      \ifx\sinstitute\empty
      \else
        {\small\sinstitute}
      \fi
    \vskip 5ex
  \endgroup
}
\makeatother
\renewenvironment{abstract}{\begin{center}
  {\large\sffamily\bfseries Abstract: }
  \begin{minipage}[t]{0.75\textwidth}
}{\end{minipage}\end{center}\vskip 10ex}

\newcommand{\myfigure}[3]{
  \begin{figure}[#1]
    \begin{center}
      #2\\
      \parbox[t]{\widthof{#2}}{\caption{#3}}
    \end{center}
  \end{figure}
}
\newcommand{\mytable}[3]{
  \begin{table}[#1]
    \begin{center}
      #2\\
      \parbox[t]{\widthof{#2}}{\caption{#3}}
    \end{center}
  \end{table}
}

\newcommand{\Sherpa}{S\protect\scalebox{0.8}{HERPA}\xspace}
\newcommand{\Comix}{C\protect\scalebox{0.8}{OMIX}\xspace}

\newcommand{\CSS}{C\protect\scalebox{0.8}{SS}\xspace}
\newcommand{\Ahadic}{A\protect\scalebox{0.8}{HADIC++}\xspace}
\newcommand{\Hadrons}{H\protect\scalebox{0.8}{ADRONS++}\xspace}
\newcommand{\Photons}{P\protect\scalebox{0.8}{HOTONS++}\xspace}

\newcommand{\Professor}{P\protect\scalebox{0.8}{ROFESSOR}\xspace}

\newcommand{\Jetphox}{J\protect\scalebox{0.8}{ET}P\protect\scalebox{0.8}{HOX}\xspace}
\newcommand{\Diphox}{D\protect\scalebox{0.8}{I}P\protect\scalebox{0.8}{HOX}\xspace}
\newcommand{\Resbos}{R\protect\scalebox{0.8}{ES}B\protect\scalebox{0.8}{OS}\xspace}
\long\def\symbolfootnote[#1]#2{\begingroup%
\def\thefootnote{\fnsymbol{footnote}}\footnote[#1]{#2}\endgroup}
\newcommand{\abs}[1]{\left| #1\right|}
\newcommand{\rbr}[1]{\left( #1\right)}
\newcommand{\abr}[1]{\langle #1\rangle}
\newcommand{\cbr}[1]{\left\{ #1\right\}}
\newcommand{\sbr}[1]{\left[ #1\right]}
\newcommand{\done}{{\rm d}}

\newcommand{\mc}[1]{\mathcal{#1}}
\newcommand{\DO}{D\O\ }
\newcommand{\dst}{\displaystyle}

\newcommand{\qcut}{Q_{\mathrm{cut}}}

\hypersetup{
  pdfauthor={Stefan Hoeche, Steffen Schumann, Frank Siegert},
  pdftitle={Hard photon production and matrix-element parton-shower merging},
  pdfkeywords={QCD, EW, Matrix Element, Parton Shower,
    Truncated Shower, CKKW}
}
\preprint{ZU-TH 19/09\\IPPP/09/94\\DCPT/09/188\\
  HD-THEP-09-28\\MCNET/09/18}
\author{Stefan H{\"o}che$^1$, Steffen Schumann$^2$, Frank Siegert$^{3,4}$}
\title{Hard photon production\\[2mm] and matrix-element parton-shower merging}
\institute{$^1$ Institut f{\"u}r Theoretische Physik, 
  Universit{\"a}t Z{\"u}rich, CH-8057 Z{\"u}rich, Switzerland\\
  $^2$ Institut f{\"u}r Theoretische Physik, 
  Universit{\"a}t Heidelberg, D-69120, Heidelberg, Germany\\
  $^3$ Institute for Particle Physics Phenomenology,
  Durham University, Durham DH1 3LE, UK\\
  $^4$ Department of Physics \& Astronomy,
  University College London, London WC13 6BT, UK\\}
\begin{document}
\maketitle
\begin{abstract}
  We present a Monte-Carlo approach to prompt-photon production, where photons
and QCD partons are treated democratically. The photon fragmentation function 
is modelled by an interleaved QCD+QED parton shower. This known technique is improved
by including higher-order real-emission matrix elements. To this end, we extend a 
recently proposed algorithm for merging matrix elements and truncated parton showers.
We exemplify the quality of the Monte-Carlo predictions by comparing them to 
measurements of the photon fragmentation function at LEP and to measurements 
of prompt photon and diphoton production from the Tevatron experiments.
\end{abstract}
\section{Introduction}
\label{sec:intro}

The measurement of final states containing photons at large transverse momenta 
plays a key role in collider experiments. Most prominently at hadron colliders 
inclusive diphoton or diphoton + jet signatures are promising channels to search for a 
light Higgs boson~\cite{Abazov:2008it,*Aaltonen:2009ga,*Aad:2009wy,*Ball:2007zza,
*Rainwater:1997dg,*Abdullin:1998er}. 
Signatures with photons might furthermore provide access to physics beyond the 
Standard Model like supersymmetry or extra spatial dimensions 
\cite{Hinchliffe:1998ys,*Giudice:1998ck,*Davoudiasl:2000wi,*Macesanu:2002ew}. 
Less spectacular but extremely important though, photon + jet final state
can be used to determine the absolute energy scale of low-$p_T$ jets~\cite{
  Bhatti:2005ai,*Abbott:1998xw,*Golutvin:2008zz} 
and to constrain the gluon distribution inside the beam 
hadron~\cite{Aurenche:1988vi,*Vogelsang:1995bg,*Fontannaz:2003yn}. The success 
of the outlined physics menue however strongly depends on our ability to thoroughly 
understand and accurately simulate such prompt-photon production processes in 
the context of the Standard Model.

In the framework of perturbation theory the mechanism of hard-photon production is twofold.
A photon can be well-separated from any other particle in the collision, 
which makes it possible to describe the reaction with fixed-order matrix elements.
The fact that these matrix elements include initial- and/or final-state QCD partons
necessitates an all-orders resummation of large logarithmic QCD corrections, which are
then absorbed into Parton Distribution Functions (PDFs) and Fragmentation Functions (FFs).
Due to its vanishing mass, a photon can also be infinitely close to an initial-
or final-state QCD parton. The related singularities in hard matrix elements are
absorbed into process-independent photon fragmentation functions~\cite{LlewellynSmith:1978dc}, 
describing the transition of a QCD parton into a bunch of hadrons and a not well-separated 
photon during the process of hadronisation. Due to the nonperturbative nature of the hadronisation process,
parton-to-photon fragmentation functions contain a nonperturbative component 
and must therefore be determined from experimental data. 
Their evolution with the factorisation scale $\mu_{F,\gamma}$ can however be calculated 
perturbatively. While the description of hard photons through matrix elements 
is said to yield the direct component of photon 
observables, the description by fragmentation functions gives the so-called fragmentation component. 
Both components are related by factorisation and must be combined to obtain a meaningful prediction 
of QCD-associated photon production.

The standard method to theoretically devise a meaningful prompt-photon cross section is 
to reflect certain experimental photon-isolation 
criteria in perturbative calculations. However, one must allow for a minimal hadronic activity 
in the vicinity of the photon. Only then it can then be ensured that all 
QCD infrared divergences are properly cancelled. 
Several such criteria are on disposal, e.g. the
cone approach \cite{Baer:1990ra,Aurenche:1989gv}, the democratic approach
\cite{Glover:1993xc,*GehrmannDeRidder:1997wx,*GehrmannDeRidder:1998ba} and the smooth 
isolation procedure \cite{Frixione:1998jh,*Frixione:1999gr}.
For both, single- and diphoton production at hadron colliders, the complete next-to-leading
order QCD corrections to respective direct and fragmentation components are known
\cite{Aurenche:1987fs,Baer:1990ra,Gordon:1994ut,Aurenche:1985yk,Bailey:1992br}.
The parton-level Monte Carlo programs \Jetphox
\cite{Catani:2002ny,*Aurenche:2006vj,*Belghobsi:2009hx} and \Diphox \cite{Binoth:1999qq,*Binoth:2000zt}
implement these NLO results numerically and allow the user to choose from different
photon-isolation criteria. The NLO corrections to the direct channel
$\gamma\gamma+1\text{jet}$ have been calculated in \cite{DelDuca:2003uz}. The results
for the loop-induced processes $gg \to \gamma\gamma$\cite{Berger:1983yi} and
$gg\to \gamma\gamma g$\cite{deFlorian:1999tp,*Balazs:1999yf,*Binoth:2003xk} are also available.
Beyond calculations at fixed order in the strong
coupling large efforts were spent on the evaluation of soft-gluon emission effects and on
the resummation of corresponding large logarithms. Soft-gluon resummation up to next-to-next-to
leading logarithmic accuracy is taken into account in the program \Resbos
\cite{Balazs:1997hv,*Balazs:1999yf,*Balazs:2006cc,*Balazs:2007hr}. The analytic result for
resumming threshold logarithms was presented in~\cite{Kidonakis:1999hq,*deFlorian:2005wf}
while small-$x$ logarithms have been studied in~\cite{Diana:2009xv}. Only recently a
first study of the prompt-photon process in the framework of Soft-Collinear Effective
Theory has been presented \cite{Becher:2009th}.

Let us note, that there is a further source of final-state photons, namely decays of 
hadrons, such as $\pi^0$ or $\eta$. However, such non-prompt production processes can to
some approximation be separated from the other two experimentally and measurements are 
usually corrected for these effects.
This is the case for all experimental data referenced in this work.

In this publication we pursue a different strategy of simulating final states including photons.
We account for the hard-production process, the QCD evolution of initial- and final-state
partons, as well as the transition of QCD partons into hadrons by
means of a multi-purpose Monte-Carlo event generator. In this context the 
lowest-order matrix elements for single- and diphoton production supplemented with
QCD parton showers correspond to the above mentioned direct component. 
The fragmentation contribution is modelled by the incorporation 
of QED effects into the parton shower. In fact, a generic algorithm to treat
photon radiation is also given by the approach of Yennie, Frautschi
and Suura~\cite{Yennie:1961ad}. This scheme is particularly suited
to compute logarithmic corrections arising from soft photon radiation,
where the coherent emission off all QED charges involved in the process
plays an important role. In this publication, however, we are primarily
interested in the production of hard, well-separated photons.
Such emissions need to be treated by an improved algorithm, see
for example~\cite{Summers:1994mc}. We therefore choose to simulate photon radiation 
using a dipole-like QED shower model. This approach only presents 
a primitive approximation to soft photon effects, but is easily realised and
no additional free parameters are introduced in the parton-shower algorithm,
cf.~\cite{Seymour:1991xa,*Seymour:1994bx}.
Similar methods are employed in most contemporary shower programs~\cite{
  Sjostrand:2006za,*Sjostrand:2007gs,*Corcella:2000bw,*Bahr:2008pv,*Lonnblad:1992tz}.
An apparent advantage is, that this method also allows for a direct
comparison with experimental data since it yields predictions at the level of the
experimentally observed particles. In particular the parton-to-photon fragmentation
functions are explicitly modelled this way. As a consequence 
no further corrections accounting for the non-perturbative parton-to-hadron transition 
need to be applied and again no additional free parameters need to be introduced.
This is crucial also for the validation of a separation of non-prompt photons
from prompt photons as mentioned above.
The {\it democratic} treatment of partons and photons in this approach combines
the direct and the fragmentation component in a very natural way.
It is well suited for comparison to experiments, where it is often necessary to study 
the impact of photon-isolation criteria.

An apparent disadvantage of the approach is that it relies on lowest-order 
matrix elements only and correspondingly higher-order QCD corrections are
taken into account in the approximation of the parton shower only.
We improve on this deficiency by including higher-order 
real-emission matrix elements. Parton-shower simulations 
supplemented with multi-leg matrix elements have become a standard tool for the 
description of QCD radiation accompanying the production of massive gauge bosons%
~\cite{Krauss:2004bs,*Krauss:2005nu,*Gleisberg:2005qq,*Alwall:2007fs} or coloured 
heavy states~\cite{Mangano:2006rw,*Alwall:2008qv,*Plehn:2008ae}. In this line we 
extend the formalism originally presented in \cite{Hoeche:2009rj} to the case of 
prompt-photon production. This process, as stated before, introduces the additional complication of a 
second source of photon production -- the fragmentation component -- which is 
not present for massive gauge bosons. While in most cases of $W$- or $Z$-boson
production the massive boson is the hardest object in the interaction, the photon
is unlikely to play this role in most prompt-photon events.
We lay out a formalism that is capable of coping with this situation and 
allows to consistently combine tree-level matrix elements of variable
photon and QCD parton multiplicity with a combined QCD+QED parton-shower model.
Photons and QCD partons are treated fully democratically in this scheme.

The outline of this paper is as follows. In Section~\ref{sec:shower} 
we introduce our method for simulating photon production using a 
dipole-like parton shower model. Section~\ref{sec:meps} presents 
the formalism for combining QCD+QED matrix-elements of different final-state 
multiplicity with the parton shower.
We also discuss how to efficiently incorporate a given
photon-isolation criterion. In Section~\ref{sec:results} we present the results 
of our Monte Carlo analysis and discuss the interplay between the direct 
photon contribution and the fragmentation component in our approach. 
Finally, Section~\ref{sec:conclusions} contains our conclusions.

\section{Interleaved QCD+QED parton evolution}
\label{sec:shower}
In this section we briefly recall a formalism which is used in our 
simulations to generate the combined QCD+QED parton evolution.
Similar approaches are implemented in most contemporary shower programs~\cite{
  Sjostrand:2006za,*Sjostrand:2007gs,*Corcella:2000bw,*Bahr:2008pv,*Lonnblad:1992tz}.
For simplicity of the argument we focus on pure final-state evolution.
Note that any parton shower algorithm is uniquely defined by three ingredients:
The first is the Sudakov form factor, $\Delta_a(\mu^2,Q^2)$, i.e.\ the probability 
for a given parton, $a$, not to radiate another parton between the two evolution 
scales $Q^2$ and $\mu^2$. The second is the ordering or evolution variable.
While the Sudakov form factor defines the anomalous-dimension matrix, and thereby
the functional form of logarithms which are resummed, the evolution variable 
selects their argument, i.e.\ it defines a ``direction'' in the phase space, 
into which the evolution is performed. The third ingredient of a parton-shower
model is the method, which is applied in order to reshuffle momenta of already 
existing partons when one of them goes off-shell to allow for a branching process.

For QCD parton evolution we choose to employ the parton-shower algorithm 
of~\cite{Schumann:2007mg}, including the ordering variables for initial-state 
evolution proposed in~\cite{Hoeche:2009rj}. This means that the Sudakov form factor
for final-state evolution reads
\begin{equation}\label{eq:sudakov_fs_qcd}
  \Delta_{(ij)}^{\,\rm QCD}(\mu_0^2,Q^2)\,=\;
    \exp\cbr{\,-\int_{\mu_0^2}^{Q^2}\frac{\done {\rm k}_\perp^2}{{\rm k}_\perp^2}
      \int_{\tilde{z}_-}^{\tilde{z}_+}\done\tilde{z}\,\sum\limits_{i,k}
      \frac{1}{2}\,{\mc K}_{(ij)i,k}^{\,\rm QCD}(\tilde{z},{\rm k}_\perp^2)\;}\;,
\end{equation}
where
\begin{align}\label{eq:kernel_fs_qcd}
  {\mc K}_{(ij)i,k}^{\,\rm QCD}(\tilde{z},{\rm k}_\perp^2)\,=&\;
  \frac{\alpha_s({\rm k}_\perp^2)}{2\pi}\;J({\rm k}_\perp^2,\tilde{z})\,
    \abr{{\rm V}_{(ij)i,k}^{\,\rm QCD}({\rm k}_\perp^2,\tilde{z})}\;
  &&\text{and}
  &\tilde{z}\,=&\;\frac{p_ip_k}{(p_i+p_j)p_k}\;.
\end{align}
The Jacobian factor $J$ and the spin-averaged dipole functions $\abr{\rm V}$ are defined 
in~\cite{Schumann:2007mg}. The sums run over all possible splitting products $i$ 
and all possible spectator partons $k$ of the splitting parton $(ij)$. 
The ordering parameter is the invariant transverse momentum squared
\begin{equation}
  {\rm k}_\perp^2\,=\;\rbr{Q^2-m_i^2-m_j^2-m_k^2}\,y_{ij,k}\,
    \tilde{z}_i(1-\tilde{z}_i)-(1-\tilde{z}_i)^2\,m_i^2-\tilde{z}_i^2\,m_j^2\;,
\end{equation}
where $Q=p_i+p_j+p_k$, $y_{ij,k}=p_ip_j/(p_ip_j+p_jp_k+p_kp_i)$ and $m$ are the parton masses.

Since QCD and QED emissions do not interfere, 
their corresponding emission probabilities factorise trivially. A combined QCD+QED 
evolution scheme is thus obtained by employing the combined Sudakov form factor
\begin{equation}\label{eq:sudakov_fs}
  \Delta(\mu_0^2,Q^2)\,=\;\Delta^{\rm QCD}(\mu_0^2,Q^2)\;\Delta^{\rm QED}(\mu_0^2,Q^2)\;,
\end{equation}
where 
\begin{equation}\label{eq:kernel_fs_qed}
  {\mc K}_{(ij)i,k}^{\,\rm QED}(\tilde{z},{\rm k}_\perp^2)\,=\;
  \frac{\alpha({\rm k}_\perp^2)}{2\pi}\,J({\rm k}_\perp^2,\tilde{z})\,
    \abr{{\rm V}_{(ij)i,k\,}^{\,\rm QED}({\rm k}_\perp^2,\tilde{z})}\;.
\end{equation}
Note that we use spin-averaged dipole functions, not only in the QCD, but also 
in the QED case. One possible improvement of the present algorithm would therefore 
be to include the spin-dependent splitting kernels. However, in the domain of hard-photon radiation that 
we are interested in, this can simply be done by employing the full real-emission 
matrix element instead. No special treatment is therefore necessary in the
parton shower.

The functional form of the spin-averaged splitting kernels is largely constrained
by the infrared singularity structure of one-loop QCD amplitudes. It is, however, 
not fixed and one has the freedom to incorporate non-singular pieces, which can help 
to improve the predictions of dipole-shower simulations, cf.~\cite{Carli:2009cg}. 
Likewise, the construction of the splitting kinematics is largely constrained by the
phase-space variables selected in the splitting. It is, however, not fixed and one
has one additional degree of freedom, which corresponds to the choice of the angular 
orientation of the splitter-spectator system with respect to the remaining particles. 
One of the most prominent criticisms of the new dipole-like 
parton-shower models is the seemingly unphysical recoil strategy employed in configurations 
with initial-state splitter and final-state spectator. This strategy is entirely due 
to a choice for the momentum mapping between leading-order and real-emission kinematics,
which was initially proposed in~\cite{Catani:1996vz,*Catani:2002hc}.
The transverse momentum of the emitted parton is thereby compensated by the spectator, 
leaving not only the virtuality of the splitter-spectator system invariant, but also 
its complete four-momentum. One can easily imagine a different recoil strategy, where 
the transverse momentum of the emission is instead compensated by the set of all 
final state particles. Such an approach was recently suggested in~\cite{Platzer:2009jq} 
for the case of massless partons. We extend it to the fully massive case in 
Appendix~\ref{sec:kinematics} and investigate the corresponding effects on the
Monte-Carlo results in Sec.~\ref{sec:results}.

It is otherwise straightforward to extend the above algorithm to initial-state showering.
The only subtlety in this context arises from the fact that the fully democratic 
approach pursued here also allows initial-state photon splitting into a quark-antiquark 
pair, with the quark (antiquark) initiating the hard scattering. In this case 
parton distributions which incorporate QED effects are in principle necessitated.
Even though such PDF fits exist (e.g.~\cite{Martin:2004dh}), the corresponding
effects on physical observables should be very small, such that the usage of PDF's
without QED contribution does not pose a conceptual problem.

An apparent disadvantage of the above algorithm for generating QED emissions using a parton shower 
is the low efficiency with which isolated photons will be produced. This problem
is dealt with in Appendix~\ref{sec:enhance}, where we introduce a method to enhance 
the corresponding emission probability, at the price of generating weighted events.

\section{Merging QCD+QED matrix elements and truncated showers}
\label{sec:meps}
In this section we discuss an algorithm for merging tree-level QCD+QED matrix elements 
and parton showers, based on the method proposed in~\cite{Hoeche:2009rj}.
We treat photons and QCD partons democratically, i.e.\ higher-order tree-level 
matrix elements can be of order $\alpha^n \alpha_s^m$ compared to the 
leading order. If $n>0$, they may contribute to an observed hard-photon
final state. In this respect, the inclusion of higher-order real corrections
corresponds to shifting the simulation of hard-photon production from the parton-shower
to the matrix-element domain.

The merging approach presented in~\cite{Hoeche:2009rj} is essentially based on replacing 
the splitting kernels of the parton shower by the appropriate ratio of full
tree-level matrix elements in the domain of hard-parton radiation. This domain 
is identified by simple phase-space slicing. The slicing parameter, the so-called
jet criterion, is given in terms of parton momenta $p_i$, $p_j$ and $p_k$ as
\begin{equation}\label{eq:jet_criterion}
  Q_{ij}^2\,=\;2\,p_i p_j\,\min\limits_{k\ne i,j}\,
    \frac{2}{C_{i,j}^k+C_{j,i}^k}\;,
  \quad\text{where}\quad
  C_{i,j}^k\,=\;\left\{\begin{array}{cc}
    {\dst \frac{p_ip_k}{(p_i+p_k)p_j}-\frac{m_i^2}{2\,p_ip_j}\,}
    &\text{if $j = g$}\\[1.5em]
    1 &\text{else}\\
    \end{array}\right.\;.
\end{equation}
For initial-state partons, one considers the splitting process
$a\to\rbr{aj}\,j$ instead of $\rbr{ij}\to i\,j$. With the momentum 
of the combined particle $\rbr{aj}$ given by $p_{aj}=p_a-p_j$, one then {\em defines}
$C_{a,j}^k\,=\;C_{\rbr{aj},\,j}^k\;$. The minimum in Eq.~\eqref{eq:jet_criterion} 
is over all possible colour partners $k$ of the combined parton $\rbr{ij}$.
The jet criterion $Q_{ij}^2$ identifies -- to leading-logarithmic accuracy --
the most likely splitting in a dipole-like parton cascade leading to the 
set of final-state momenta $\cbr{p}$.
The phase-space slicing is now implemented by selecting a cut value $Q_{\rm cut}$
and defining the evolution kernels ${\mc K}^{\,\rm ME}$ and ${\mc K}^{\,\rm PS}$
for matrix-element and parton-shower domain as
\begin{align}\label{eq:kernel_meps}
  \mc{K}^{\rm ME}_{(ij)i,k}(\tilde{z},{\rm k}_\perp^2)=&\;\mc{K}_{(ij)i,k}(\tilde{z},{\rm k}_\perp^2)\;
    \Theta\sbr{\vphantom{\sum}Q_{ij}(\tilde{z},{\rm k}_\perp^2)-Q_{\rm cut}}\;,\\
  \mc{K}^{\rm PS}_{(ij)i,k}(\tilde{z},{\rm k}_\perp^2)=&\;\mc{K}_{(ij)i,k}(\tilde{z},{\rm k}_\perp^2)\;
    \Theta\sbr{\vphantom{\sum}Q_{\rm cut}-Q_{ij}(\tilde{z},{\rm k}_\perp^2)}\;.
\end{align}
The kernel ${\mc K}^{\rm ME}$ is then replaced by appropriate ratios of
tree-level matrix elements up to a given maximum multiplicity.
A detailed description of the corresponding algorithm can be found
in~\cite{Hoeche:2009rj}. It is obvious that the same procedure can 
be applied to QED emissions, once they are resummed by the parton
shower using Eq.~\eqref{eq:sudakov_fs}. It is then in principle
possible to define two separate slicing cuts, $Q_{\rm cut}^{\,\rm QCD}$
and $Q_{\rm cut}^{\,\rm QED}$, which account for the merging of QCD
and QED tree-level matrix elements with the parton shower, respectively.
Within the context of this work, we choose to leave these slicing cuts
identical, since the typical ``hardness'' of a hard well-separated 
final-state photon is similar to the one of a final-state QCD jet.

In prompt photon production processes we might be confronted with a situation 
which cannot arise in pure QCD events, namely that a single, perturbatively 
produced particle -- the photon -- is identified out of potentially many 
particles forming a broad jet. Several methods exist to achieve this identification.
In the democratic approach~\cite{Glover:1993xc,*GehrmannDeRidder:1997wx,
  *GehrmannDeRidder:1998ba} final state particles are clustered into jets, 
treating photons and hadrons equally. The obtained object is called a photon 
or a photon jet, if the energy fraction $z=E_\gamma/(E_\gamma+E_{had})$
of an observed photon inside the jet is larger than an experimentally defined 
value $z_{cut}$. In the cone approach~\cite{Baer:1990ra,Aurenche:1989gv} 
photons are required to have a minimum transverse momentum and to be isolated 
from any significant hadronic activity within a cone in $\eta$-$\phi$-space. 
Minimal hadronic activity in the vicinity of the photon (adding of the order of a 
few GeV to the total transverse momentum in the cone) must thereby be admitted to 
ensure the infrared finiteness of observables.

While the jet criterion Eq.~\eqref{eq:jet_criterion} works very well also for
photons defined by the democratic approach, in the case of the cone approach it
might not be appropriate to separate matrix-element and parton-shower domain.
Note that the main idea of the merging procedure is to improve the parton-shower
prediction with fixed-order matrix elements in those regions of phase space
which are relevant for the analysis of multi-jet (multi-photon) topologies.
In this respect, it is certainly desirable that experimental requirements 
are reflected by $Q_{ij}^2$. This is possible because the jet criterion, 
Eq.~\eqref{eq:jet_criterion}, is not fixed, but rather chosen conveniently 
to reflect the singularity structure of next-to-leading order real-emission 
amplitudes in QCD~\cite{Hoeche:2009rj}.
Moreover it is a flavour-dependent measure, which allows to redefine it just for 
branching processes involving photons. The most common experimental requirements 
of a minimum transverse momentum and an isolation cone in $\eta$-$\phi$-space
could for example be reflected by
\begin{align}\label{eq:jet_criterion_photons}
  Q_{ij}^2\,&=\;\min\cbr{p_{\perp,i}^2,p_{\perp,j}^2}\frac{\Delta\eta_{ij}^2+\Delta\phi_{ij}^2}{D^2}
  &&\text{and}
  &Q_{ib}^2\,&=\;p_{\perp,i}^2\;,
\end{align}
where the first equation applies to final-state photons and charged 
final state particles, while the second applies to photons and charged beams.
Note that Eq.~\eqref{eq:jet_criterion_photons} is essentially equivalent to a 
longitudinally invariant jet measure~\cite{Catani:1992zp,*Catani:1993hr}. 
One can now increase the ratio of photons produced through matrix elements 
over photons produced in the shower by simply lowering the value of $Q_{\rm cut}^\mathrm{QED}$.
A convenient way to obtain the largest fraction of events from hard matrix elements
is to require a jet separation below the experimental cut on the photon transverse 
momentum and by setting $D$ lower than the radius of the experimentally imposed 
isolation cone.

\section{Results}
\label{sec:results}
In this section we apply the event-generation techniques introduced above
to prompt-photon production at lepton and hadron colliders. In 
Sec.~\ref{sec:results:ee} we study the capability of our proposed QCD+QED shower
algorithm to reproduce the scale dependent photon fragmentation function measured 
in hadronic $Z^0$ decays at LEP. In Sec.~\ref{sec:results:tev1} and 
Sec.~\ref{sec:results:tev2} we turn the discussion on single- and diphoton final 
states at hadron colliders, respectively. Besides quantifying the size of the different core-process 
components in the democratic shower approach we elaborate on the impact of real-emission 
matrix-element corrections incorporated using the formalism described in Sec.~\ref{sec:meps}.
All results presented here are obtained using the Monte-Carlo event generator 
\Sherpa\cite{Gleisberg:2003xi,*Gleisberg:2008ta} in a setup described in 
Appendix~\ref{sec:sherpa}.

\subsection{The photon fragmentation function}
\label{sec:results:ee}

A crucial benchmark for the combined QCD+QED shower algorithm introduced in 
Sec.~\ref{sec:shower} is posed by the requirement to reproduce the scale-dependent 
photon fragmentation function $D_\gamma(z_\gamma,y_{\rm cut})$~\cite{Glover:1993xc},
where $z_\gamma$ is the photon's energy fraction with respect to its containing jet 
and $y_{\rm cut}$ a resolution scale, given e.g. in the Durham scheme. This 
observable was measured to very high precision in hadronic $Z^0$ decays by the 
ALEPH collaboration~\cite{Buskulic:1995au}. In this analysis events are selected 
where all final-state particles are democratically taken into account for jet 
finding. The events are subdivided into $2$-jet, $3$-jet and $\geq 4$-jet
topologies with at least one reconstructed jet containing a photon where the photon 
carries at least $70\%$ of the jet energy ($z_\gamma>0.7$) and $E_\gamma>5\;{\rm GeV}$. 
The resolution measure $y_{\rm cut}$ is varied between $0.01$ and $0.33$. The measured 
data is statistically corrected for residual hadronic decay backgrounds and initial-state
radiation off the incoming leptons.

\begin{figure}[p]
\begin{centering}
  \includegraphics[width=0.38\linewidth]{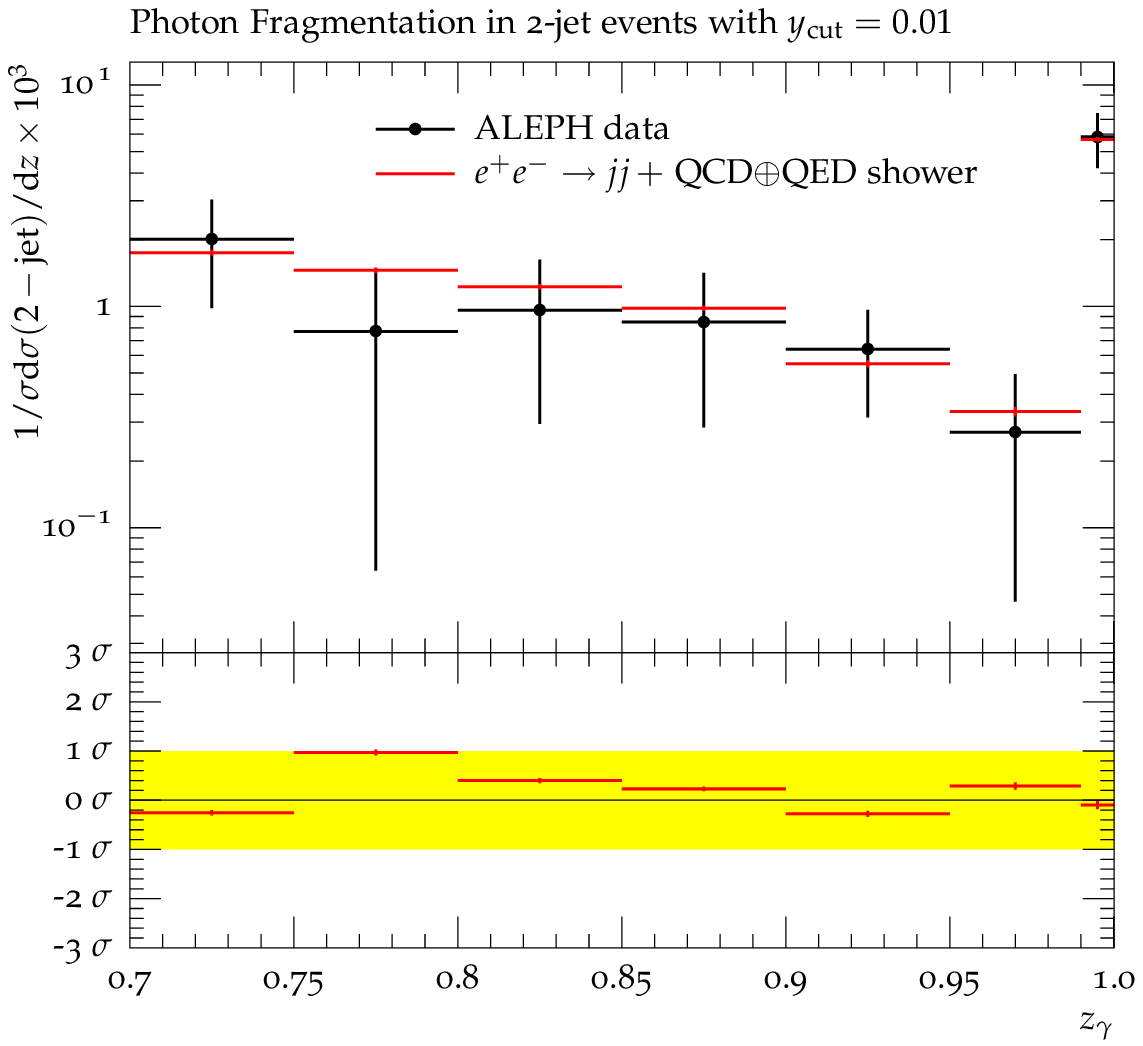}\hspace*{5mm}
  \includegraphics[width=0.38\linewidth]{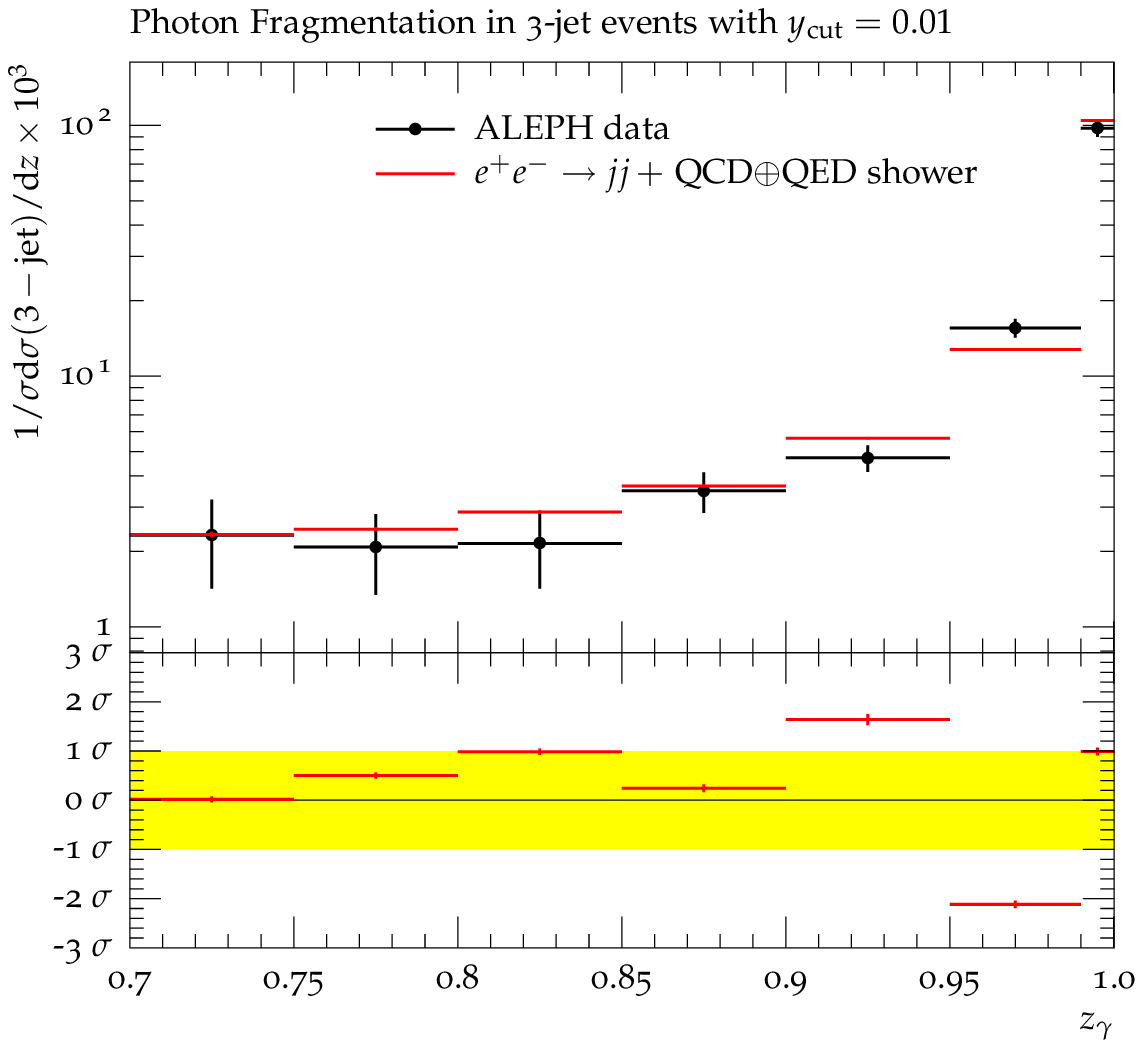}\hspace*{5mm}\\
  \includegraphics[width=0.38\linewidth]{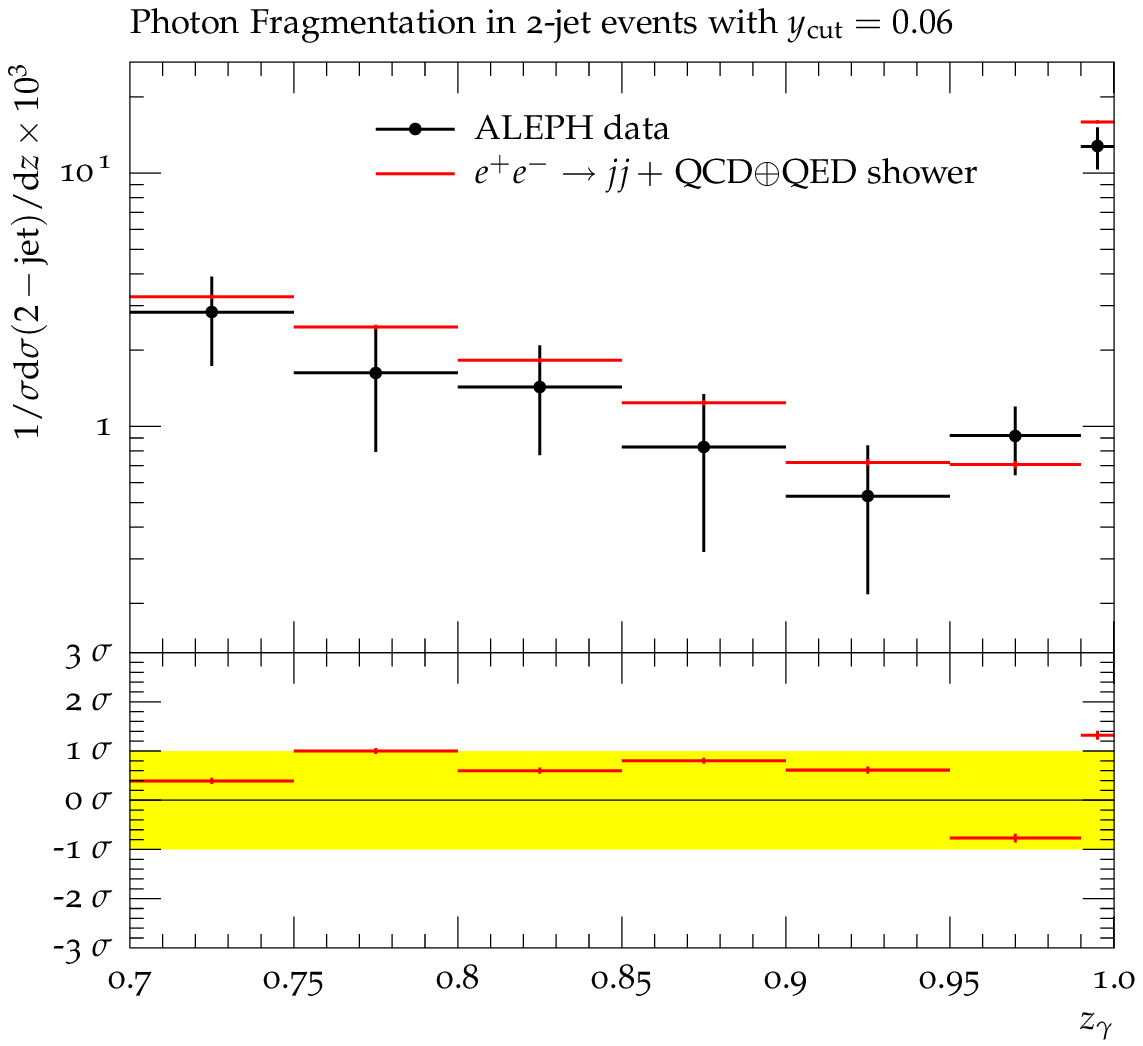}\hspace*{5mm}
  \includegraphics[width=0.38\linewidth]{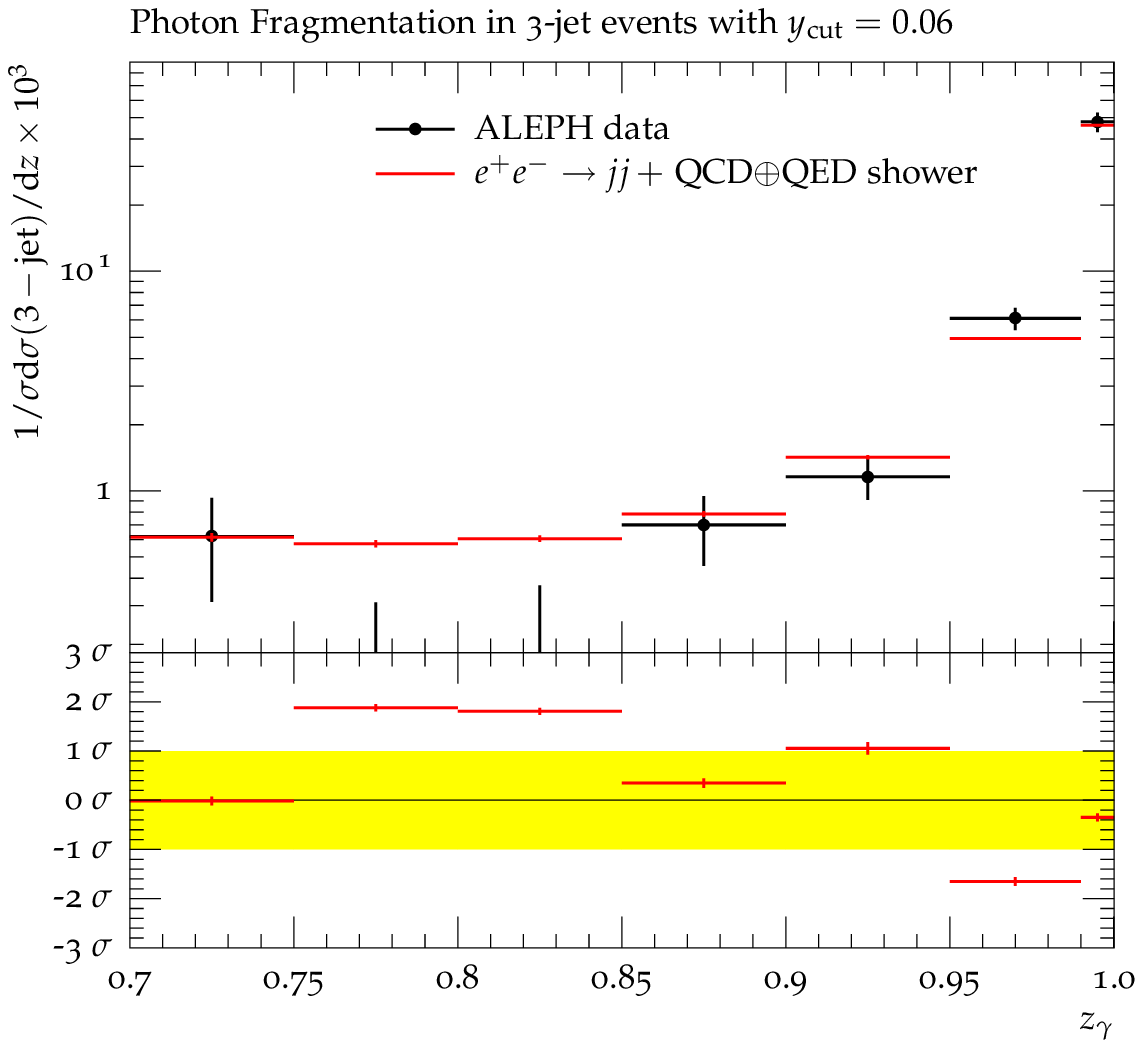}\hspace*{5mm}\\
  \includegraphics[width=0.38\linewidth]{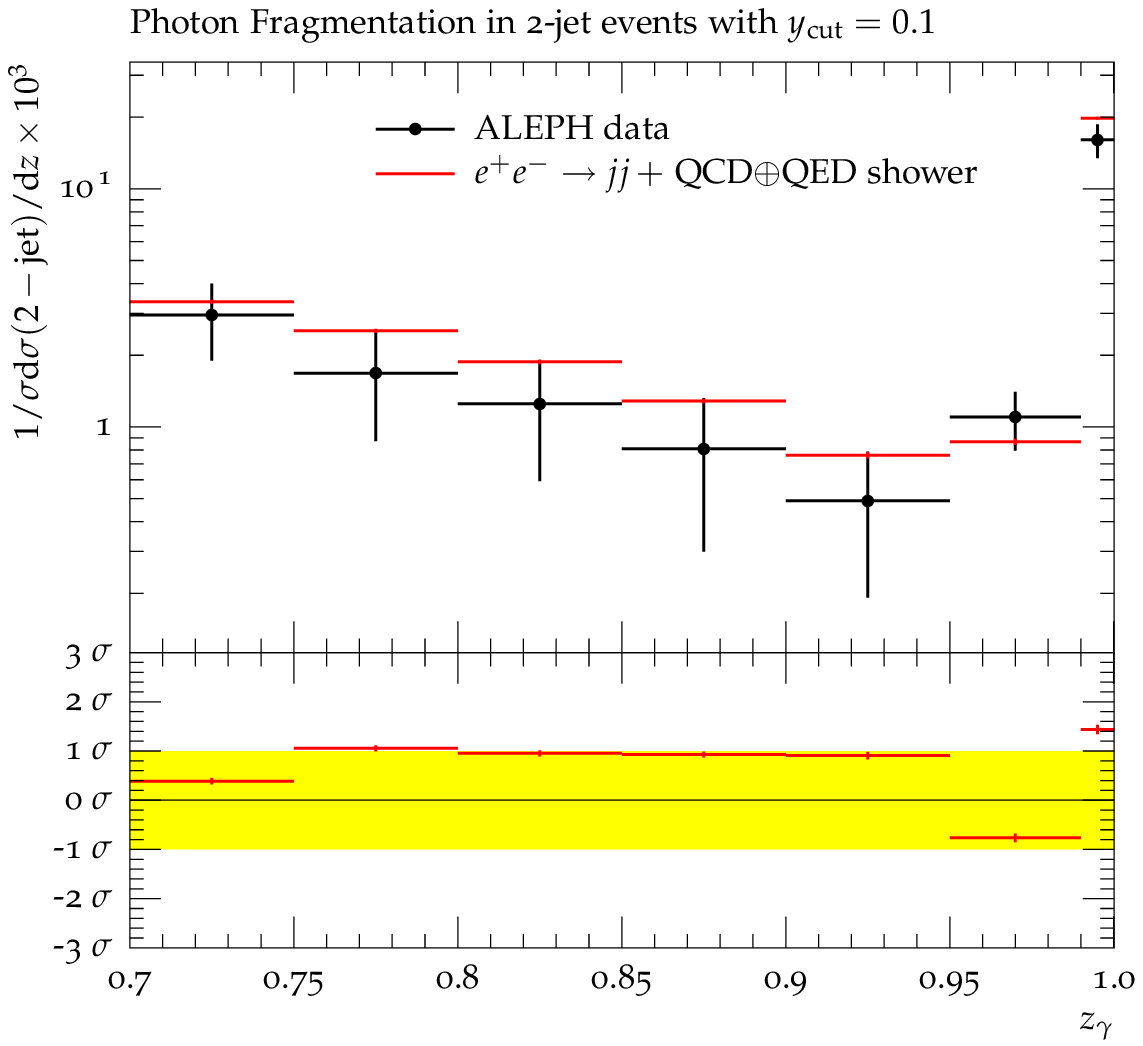}\hspace*{5mm}
  \includegraphics[width=0.38\linewidth]{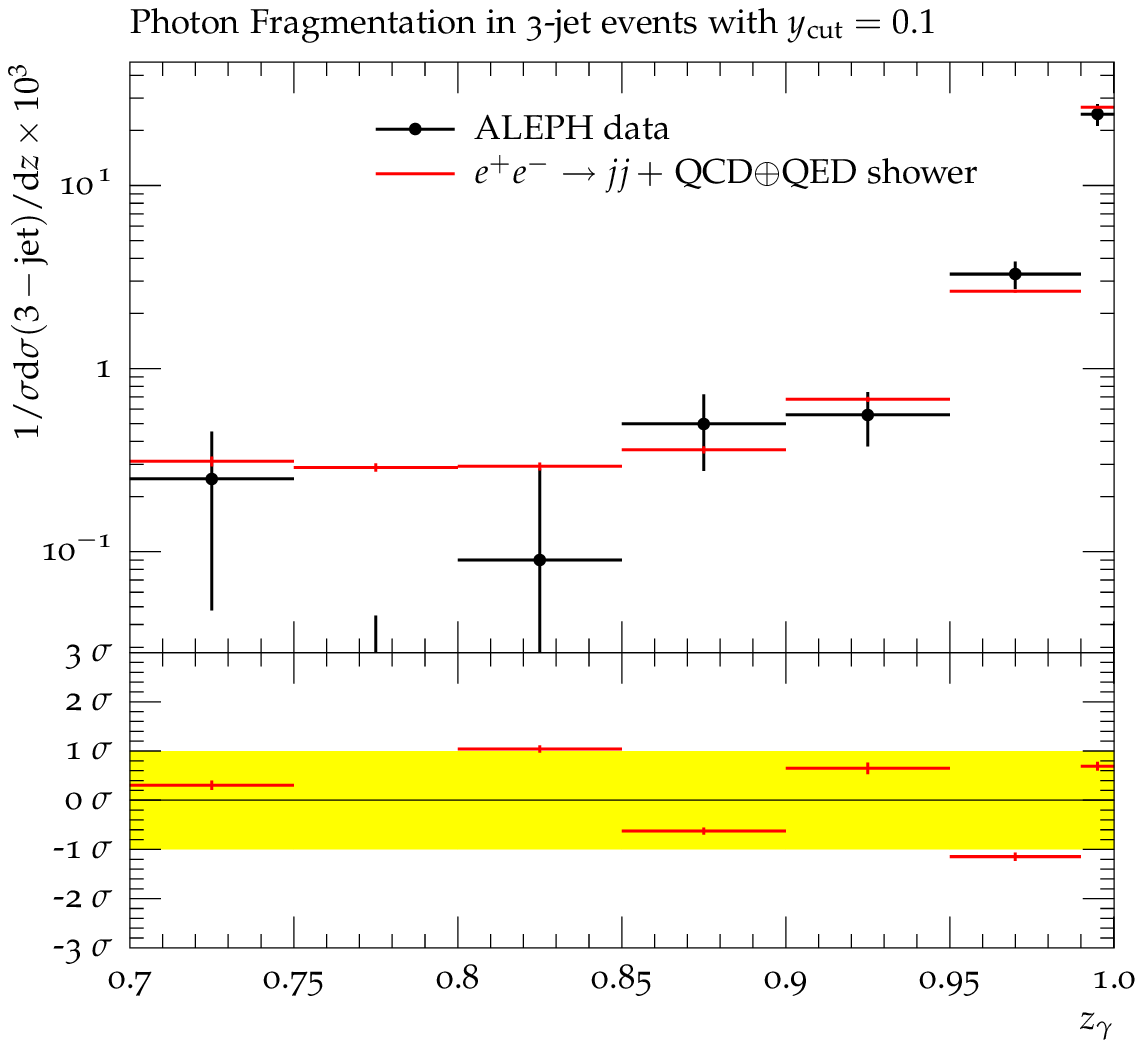}\hspace*{5mm}\\
  \includegraphics[width=0.38\linewidth]{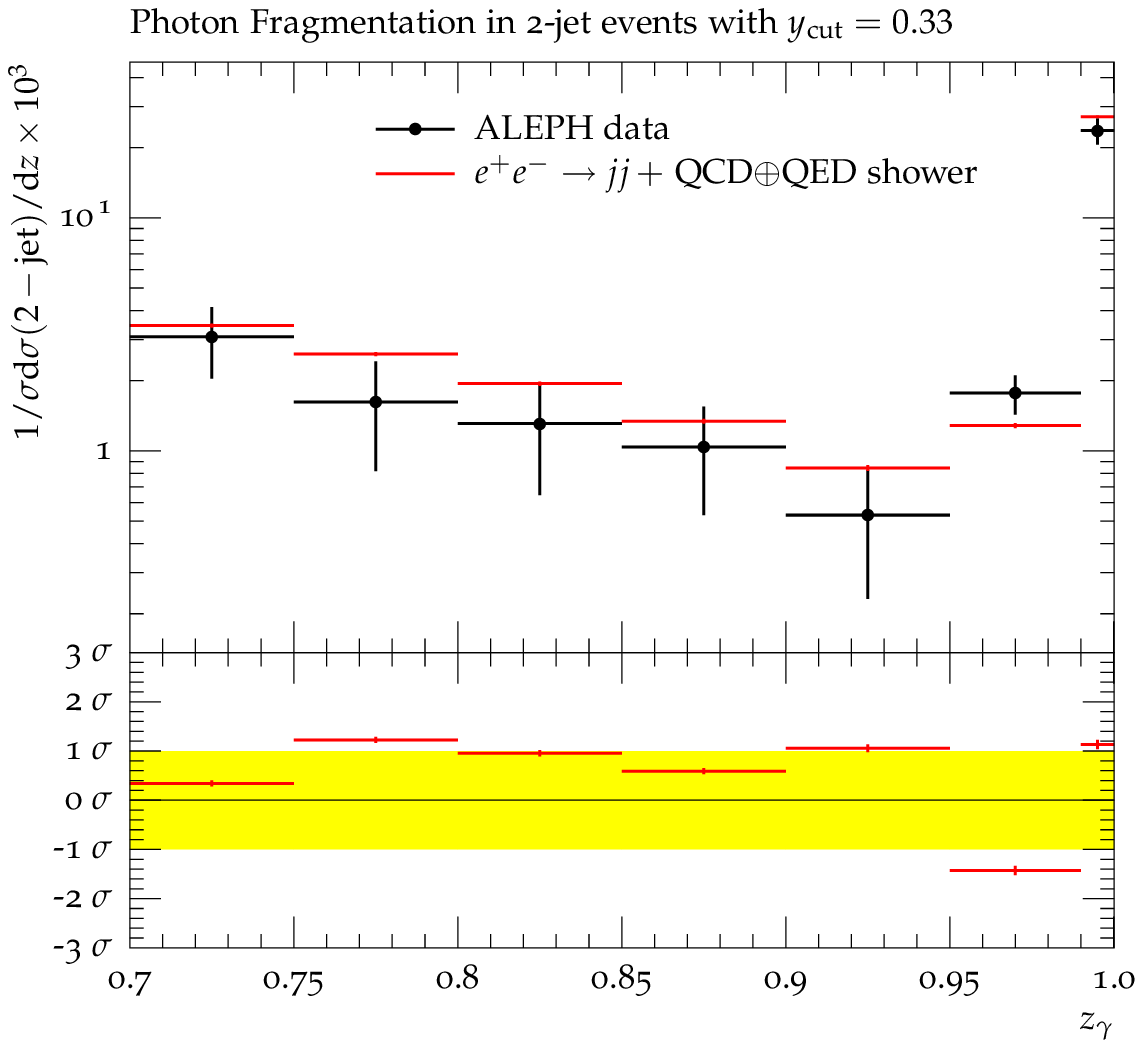}\hspace*{5mm}
  \includegraphics[width=0.38\linewidth]{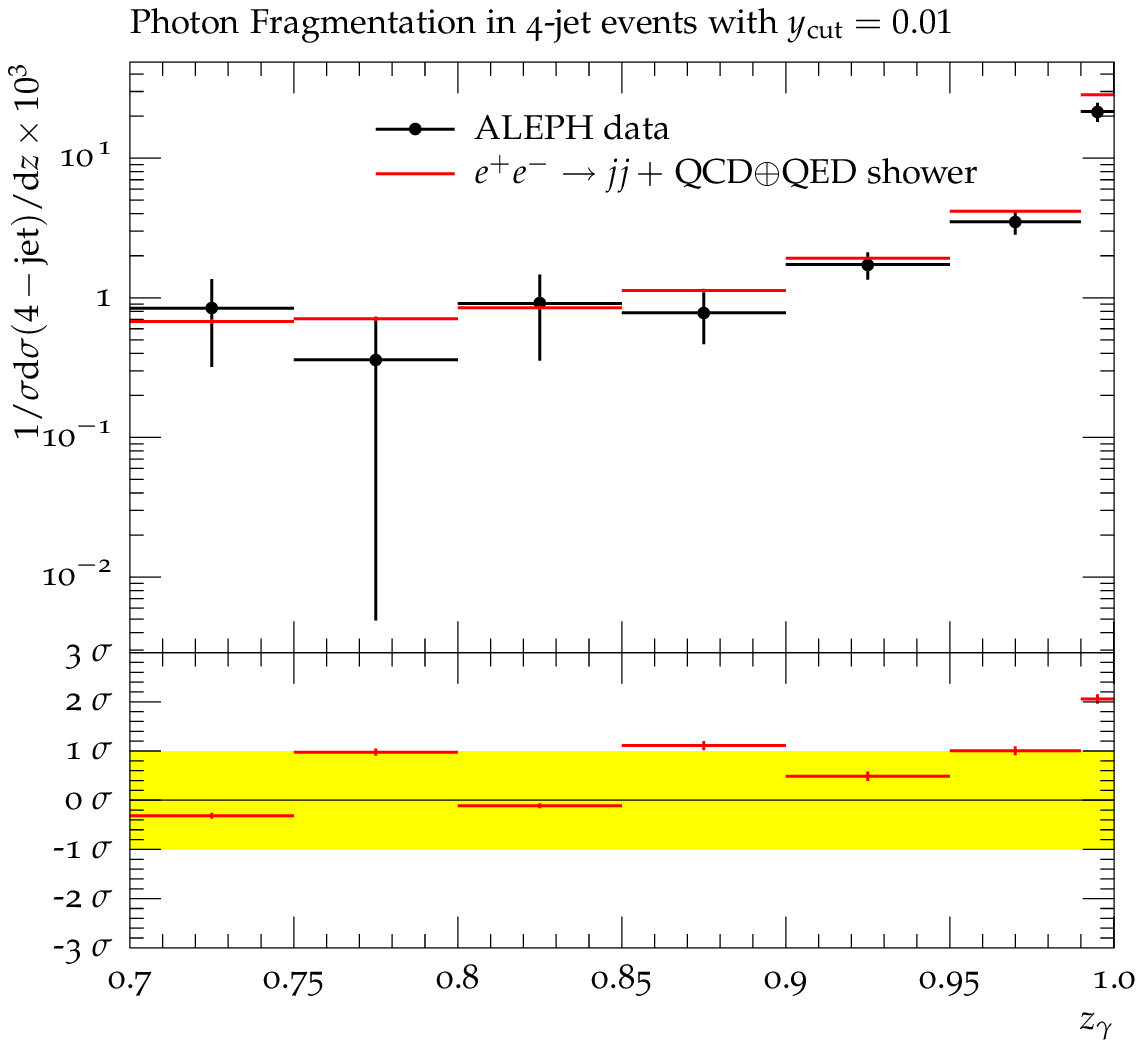}
  \caption{The $z_\gamma$ distribution measured in hadronic $Z^0$ decays 
by ALEPH~\protect\cite{Buskulic:1995au} for $2$-jet, $3$-jet and $\geq 4$-jet 
events at different Durham resolution $y_{\rm cut}$. The theory result corresponds to QCD+QED 
shower evolution of the leading-order $q\bar q$ process, taking into account hadronisation corrections.}
  \label{fig:eeshower}
\end{centering}
\end{figure}

Figure~\ref{fig:eeshower} shows a comparison between our hadron-level Monte-Carlo 
results and the data from~\cite{Buskulic:1995au}. In the left column the $z_\gamma$ 
distribution for $2$-jet events at four different $y_{\rm cut}$ values, namely 
$0.01$, $0.06$, $0.1$ and $0.33$, are shown. The right column shows corresponding 
results for $3$-jet events at $y_{\rm cut}=0.01,\,0.06,\,0.1$ and $\geq 4$-jet events 
at $y_{\rm cut}=0.01$. For all the data $z_\gamma=1$ corresponds to completely 
isolated photons, which is reflected by a strong peak in the $z_\gamma$ distribution. 
At the parton level the lowest-order contribution to fully 
isolated photon production corresponds to $q\bar q \gamma$ final states
where the quarks form one jet and the photon makes up the other one. At the
hadron level however this sharp peak gets somewhat broadened by hadronisation 
effects due to the association of soft hadrons with the photon by the jet-clustering 
algorithm. Our Monte-Carlo simulation agrees very well with the data for the
measured $z_\gamma$ range in the three topology classes at the given jet resolutions.
This can be seen as a strong indication that the proposed QCD+QED shower scheme 
is indeed appropriate to describe hard photon radiation. 

The above analysis is especially tailored to study the fragmentation 
component of the prompt photon production mechanism. The key point in this 
respect is the application of a democratic jet-clustering procedure which is in 
one-to-one correspondence with the democratic approach used to compute the 
photon-production rate from theory~\cite{Glover:1993xc}. It is therefore not 
obvious, that the democratic approach, also underlying our simulation, performs 
well in experimental situations where the photon must pass a strict isolation criterion, 
as discussed in Sec.~\ref{sec:meps}. We will now turn to investigate such isolated photon
final-states in more detail, focusing on their emergence at hadron colliders.

\subsection{Prompt-photon hadroproduction}
\label{sec:results:tev1}
The inclusive production of isolated photons has been measured over a wide range of 
photon transverse energies by the CDF and \DO experiments at the Fermilab Tevatron 
at $\sqrt{s}=1.96\;{\rm GeV}$. In \cite{Aaltonen:2009ty} CDF has presented a measurement 
covering $|\eta_\gamma|<1.0$ and transverse energies between 
$30 < E_{T,\gamma} < 400\;{\rm GeV}$. The photon isolation criterion used corresponds to 
the requirement that the additional transverse energy found in a cone 
of $R=\sqrt{(\Delta\phi)^2+(\Delta\eta)^2}=0.4$ around the photon is less than $2\;{\rm GeV}$.
A similar \DO measurement was described in \cite{Abazov:2005wc}. It covers photons of 
transverse momentum $p_{T,\gamma}>23\;{\rm GeV}$ up to $p_{T,\gamma}=300\;{\rm GeV}$ and 
$|\eta_\gamma|<0.9$. Photon isolation is implemented by demanding 
$(E_{R=0.4}-E_{R=0.2})/E_{R=0.2}<0.1$, where $E_{R}$ is the total energy found in a cone 
of size $R$ around the photon. Both measurements have been corrected to particle level and 
the dominant background of photon production from hadron decays, such as 
$\pi^0 \to \gamma \gamma$ and $\eta \to \gamma \gamma$, has been subtracted. We therefore 
attempt a comparison with Monte-Carlo predictions at the parton level after jet evolution.

\begin{figure}[p]
\begin{centering}
  \includegraphics[width=0.48\linewidth]{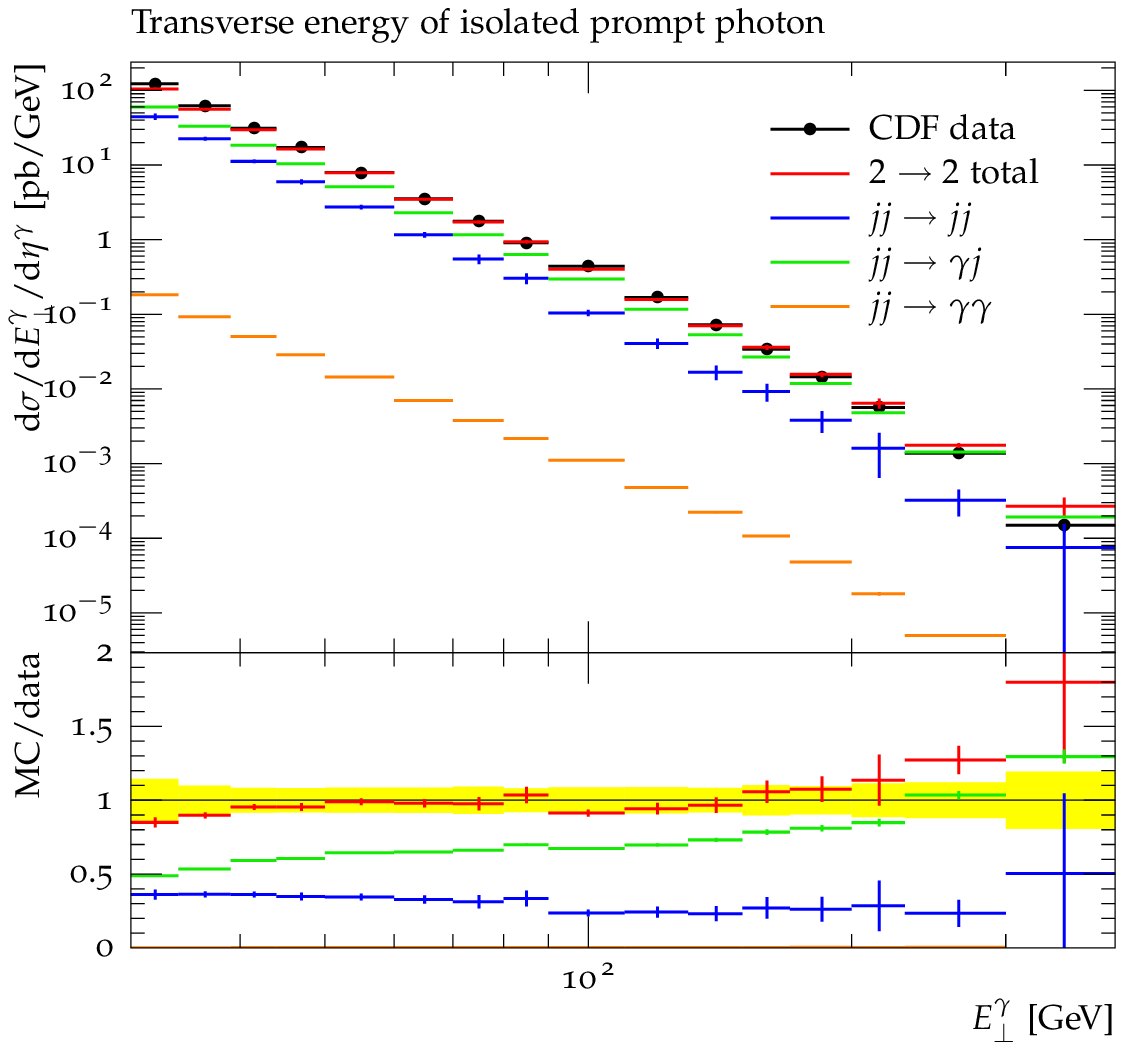}\hspace*{5mm}
  \includegraphics[width=0.48\linewidth]{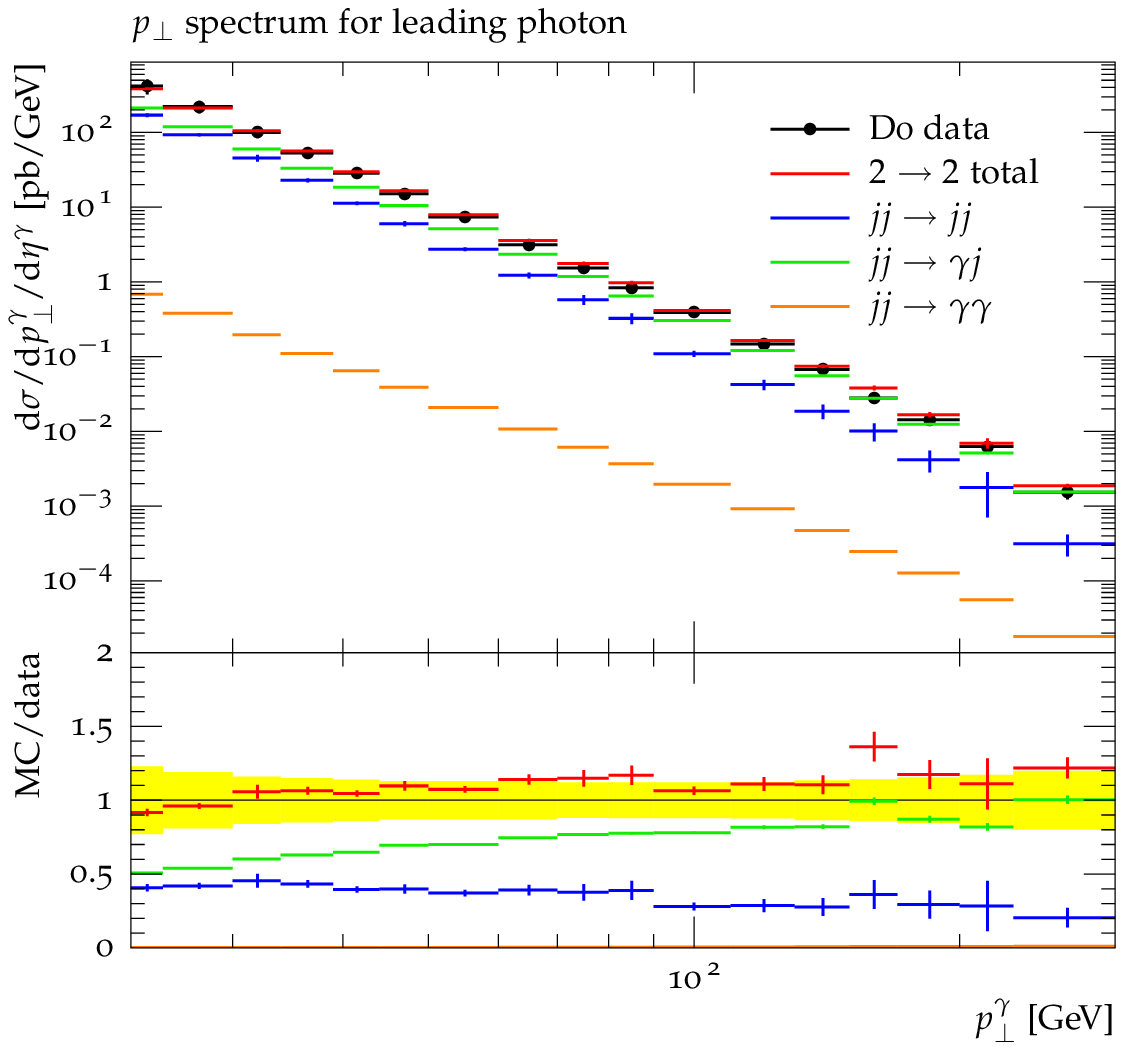}
  \caption{Inclusive photon transverse energy distribution compared to data from CDF~\protect\cite{Aaltonen:2009ty} (left) 
    and \DO\protect\cite{Abazov:2005wc} (right). The contributions of different classes of leading-order core processes are also displayed. For the notation used cf.\ the main text.}
  \label{fig:tev_pTy_conts}
\end{centering}
\end{figure}

Figure~\ref{fig:tev_pTy_conts} compares the data for 
$d^2\sigma/(dE_{T,\gamma}d\eta_\gamma)$ from \cite{Aaltonen:2009ty}, respectively 
$d^2\sigma/(dp_{T,\gamma}d\eta_\gamma)$ from \cite{Abazov:2005wc}, to our parton-level 
Monte-Carlo results, obtained using leading-order matrix elements in the democratic approach 
combined with QCD+QED shower evolution. In addition to the total result (red histograms), 
we display contributions from the different classes of partonic core processes, i.e.\ from 
dijet production ($jj\to jj$), single-photon production ($jj\to \gamma j$) and diphoton 
production ($jj\to \gamma\gamma$). Taking into account the uncertainties of the measurements 
and the finite Monte-Carlo statistics in the high-$E_{T,\gamma}$ bins our calculation 
agrees well with the data. For the CDF measurement the data has a somewhat steeper slope at 
low $E_{T,\gamma}$ and the Monte-Carlo calculation pronounces the high $E_{T,\gamma}$ end
of the spectrum. Regarding the different sources of final-state photons in our
theoretical model, the main contribution to this observable stems from single photon
production. But even though strict isolation criteria are applied, there is a
considerable fraction of dijet events, where a hard, isolated photon
is produced during the parton-shower evolution in both data 
samples. This substantiates the argument that the combined shower scheme is crucial for a 
proper description of such photon final states. The diphoton core process on the other hand 
is negligible here. 

\begin{figure}[p]
\begin{centering}
  \includegraphics[width=0.48\linewidth]{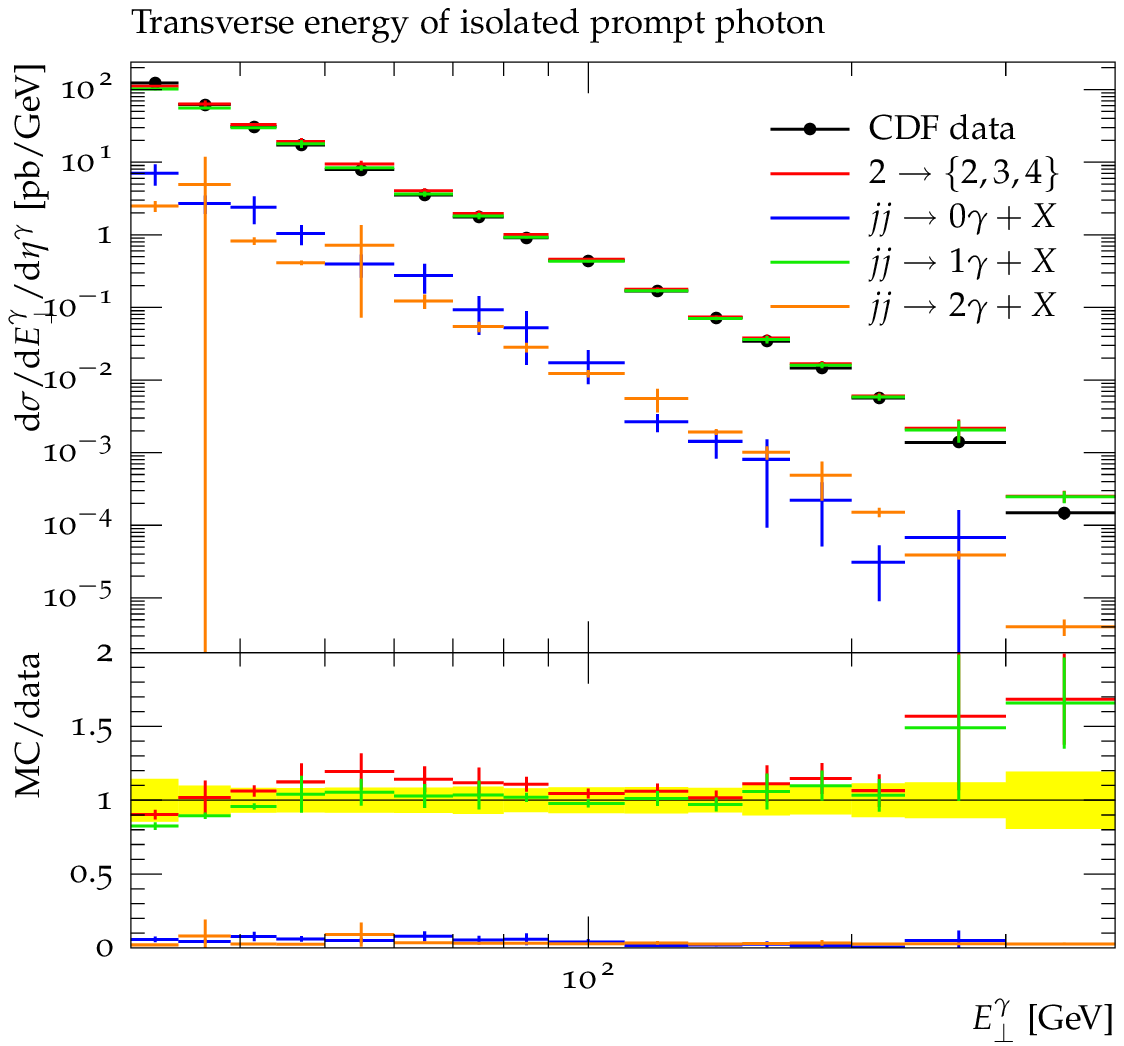}\hspace*{5mm}
  \includegraphics[width=0.48\linewidth]{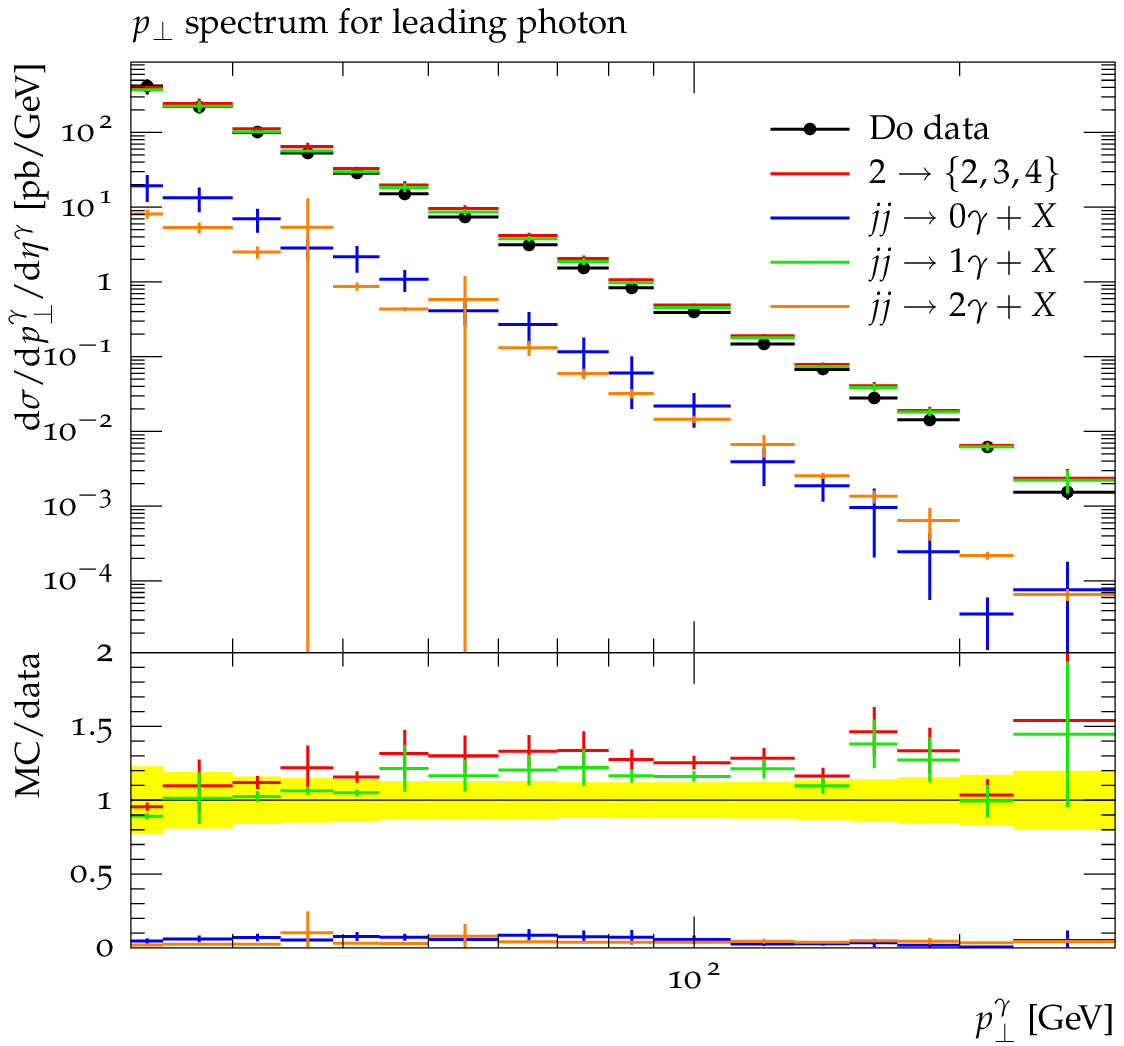}
  \caption{The inclusive photon transverse energy obtained with a QCD+QED shower simulation supplemented by 
    real-emission matrix elements with up to two additional QCD partons or photons (denoted $2\to {2,3,4}$)
    is compared to data from CDF~\protect\cite{Aaltonen:2009ty} (left) 
    and \DO~\protect\cite{Abazov:2005wc} (right). }
  \label{fig:tev_pTy_njets}
\end{centering}
\end{figure}

We now turn to study the impact of higher-order real-emission matrix elements on the results.
Therefore we supplement the pure parton-shower evolution by tree-level matrix elements with up to
two additional light QCD partons or photons using the matrix-element parton-shower merging formalism 
described in Sec.~\ref{sec:meps}. The comparison with measurements from CDF and \DO is shown in 
Fig.~\ref{fig:tev_pTy_njets}. Besides the total results (red histograms) we again present the 
sub-contributions assigned to matrix-element core processes with exclusively $0,1$ and $2$ photons 
plus a variable number of additional QCD partons. When comparing to Fig.~\ref{fig:tev_pTy_conts},
where the pure shower result is shown, it is apparent that the majority of events with 
a dijet core process in the shower simulation is now ascribed to matrix-element cores 
with one or two photons plus additional QCD partons. Thereby, what might traditionally be called 
{\it fragmentation} component gets significantly reduced and turned into what is called {\it direct}
component. For this very inclusive measurement we observe no strong variation of the total result
due to the inclusion of real-emission matrix elements. The biggest effect is a somewhat larger 
inclusive rate for the merged samples $\lesssim 20\%$. The shape of the distributions is preserved. 
This in fact has to be understood as a highly non-trivial consistency check of our merging formalism. 

\subsection{Prompt-diphoton production}
\label{sec:results:tev2}
An interesting further testbed for the democratic merging approach is diphoton production 
at hadron colliders. The CDF collaboration has measured properties of the corresponding 
final states in some detail. The analysis presented in~\cite{Acosta:2004sn}
selects leading/subleading photons with transverse momenta larger than 14/13 GeV. Those must be 
isolated from any significant hadronic activity within a distance of $R=0.4$,
by requiring the hadronic activity within this cone to yield $E_T<1$ GeV.
For the selected events the invariant mass and transverse momentum of the 
photon pair are analysed as well as the azimuthal separation between the 
photons.

It is worth noting that our MC simulations include the loop-induced
contribution $gg \to \gamma \gamma$. It has been shown~\cite{Acosta:2004sn}
that its main influence is seen in the invariant mass spectrum around $30$ GeV
where it accounts for a significant enhancement of the cross section.

\begin{figure}[t]
\begin{centering}
  \includegraphics[width=0.48\linewidth]{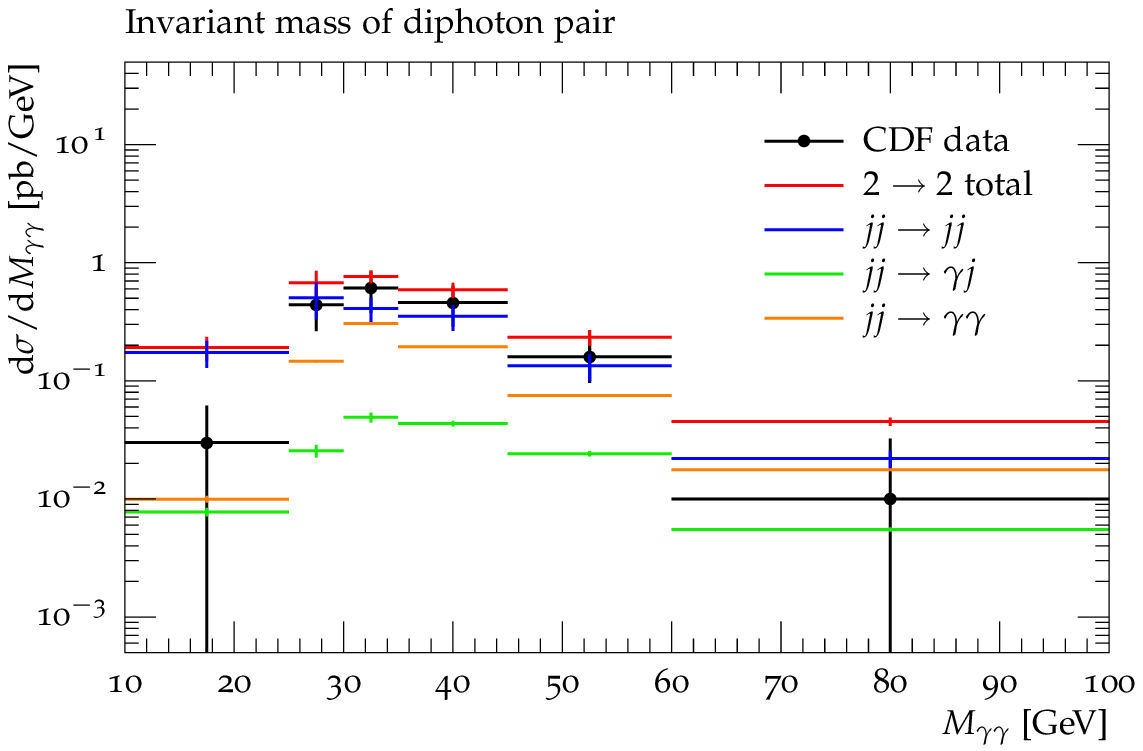}\hspace*{5mm}
  \includegraphics[width=0.48\linewidth]{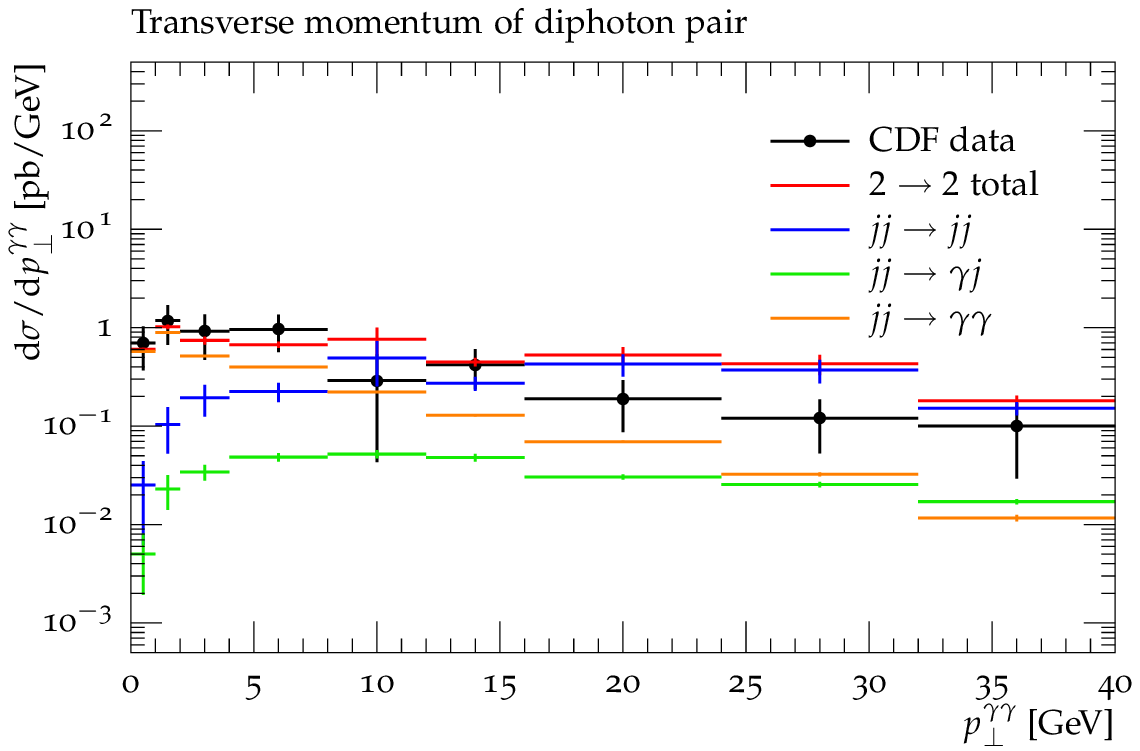}
  \caption{Properties of diphoton events measured by CDF~\protect\cite{Acosta:2004sn}.
    Displayed are the sub-contributions from different leading-order matrix
    elements and their sum.}
  \label{fig:tev_diphoton_conts}
\end{centering}
\end{figure}

To again exemplify the importance of the fragmentation contribution even for
the required isolated photons in this analysis, 
Figure~\ref{fig:tev_diphoton_conts} compares our Monte-Carlo prediction for
leading-order matrix elements plus shower evolution.
It displays the contributions from the previously introduced classes of
matrix elements (i.e.\ $jj\to jj$, $jj\to \gamma j$ and $jj\to\gamma\gamma$). 
It is evident, that the democratic treatment of photons and QCD partons is 
absolutely mandatory to describe these observables. 

From a theoretical perspective this reaction is interesting because the diphoton system 
does not have a transverse momentum when described by leading-order matrix elements. 
Hence its transverse-momentum spectrum depends strongly on the proper 
inclusion of higher-order effects. In addition, the azimuthal angle
gives a measure for the correlation of the two photons which is also
sensitive to higher-order corrections.
Especially in the region of large transverse momenta or large decorrelation
one expects these corrections to be better described by matrix elements than by
the parton shower.

\begin{figure}[p]
\begin{centering}
  \subfloat[][]{\parbox{0.46\textwidth}{
    \includegraphics[width=\linewidth]{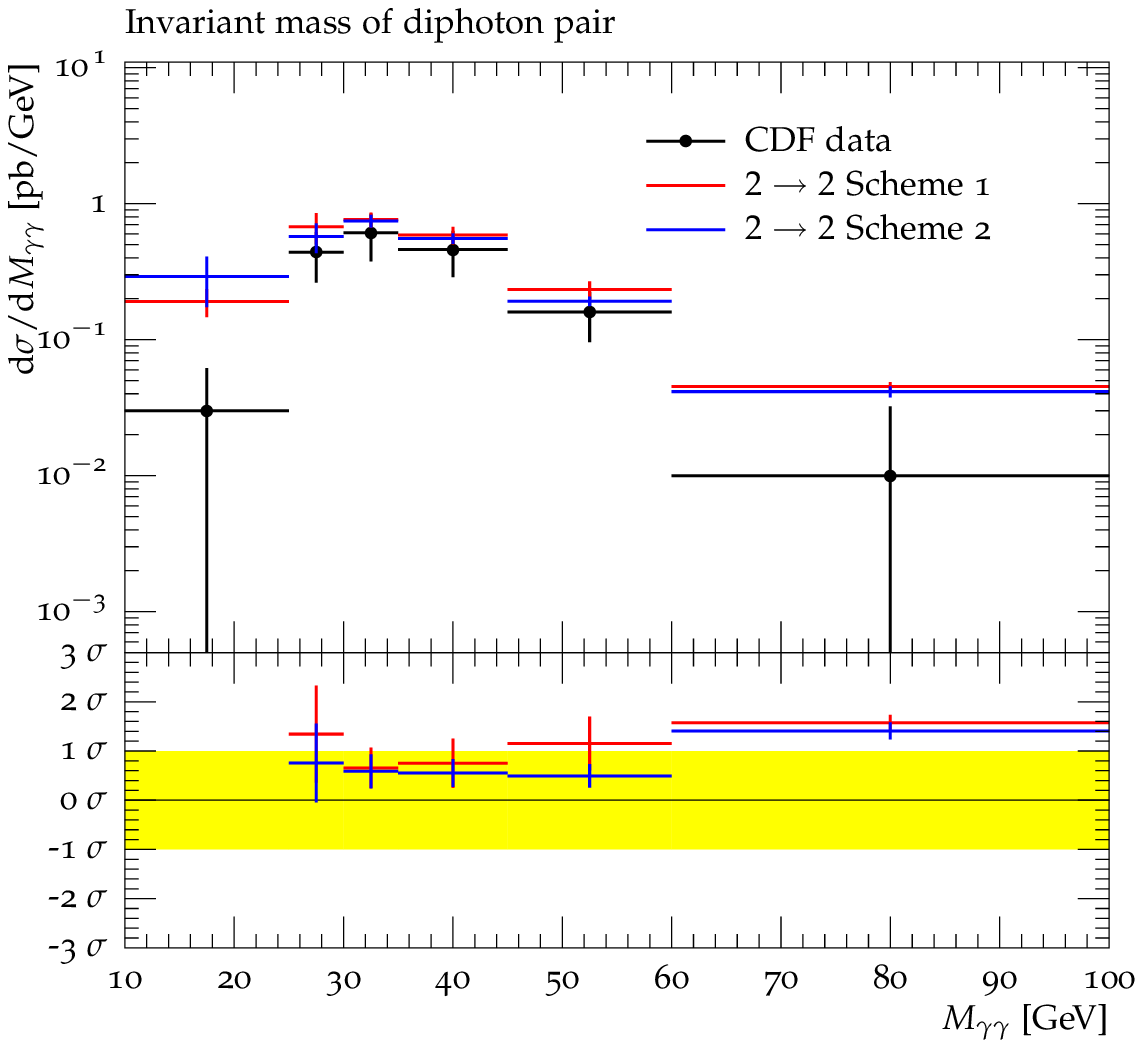}\\
    \includegraphics[width=\linewidth]{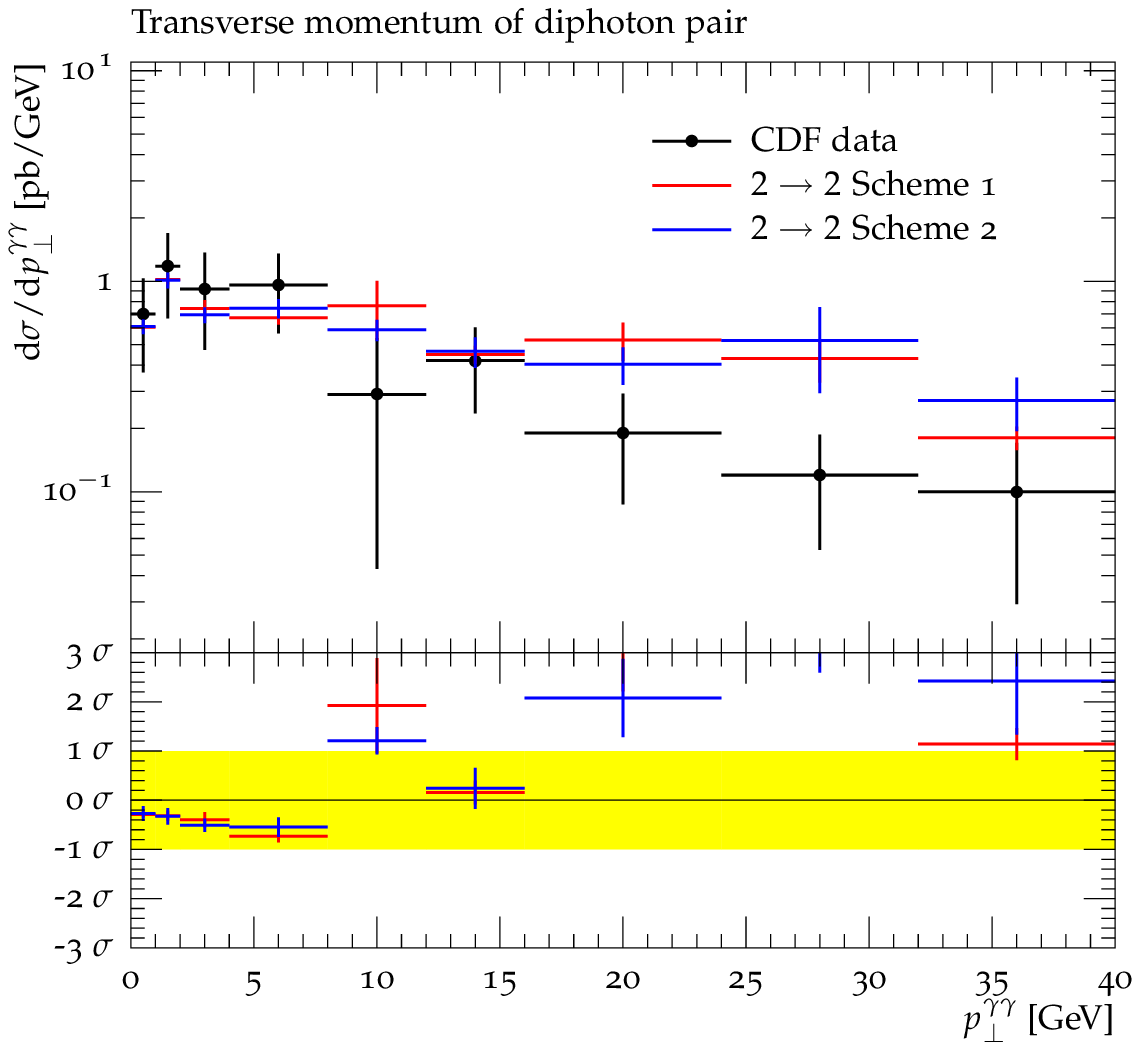}\\
    \includegraphics[width=\linewidth]{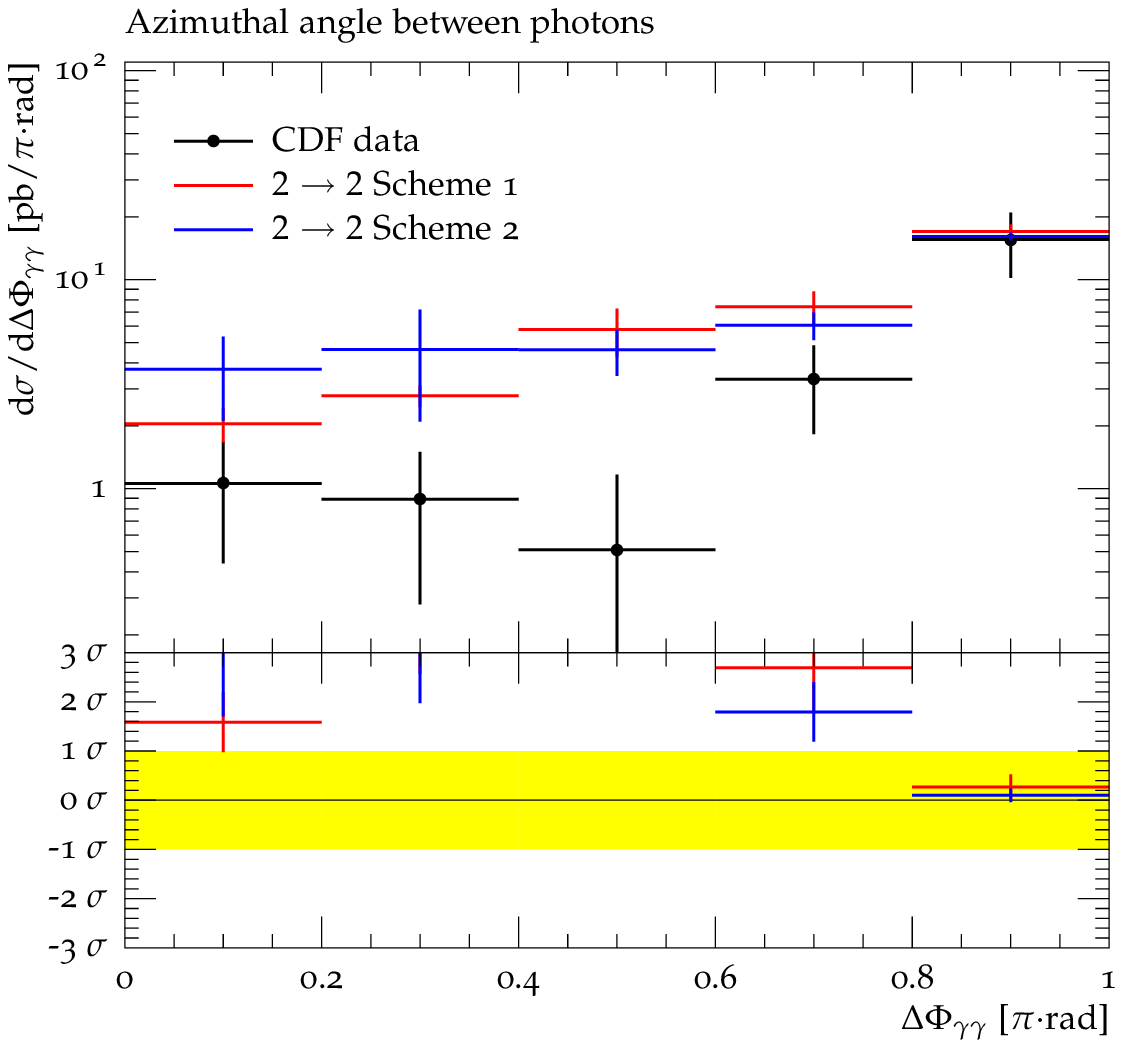}}
    \label{fig:tev_diphoton_ps}}
  \subfloat[][]{\parbox{0.46\textwidth}{
    \includegraphics[width=\linewidth]{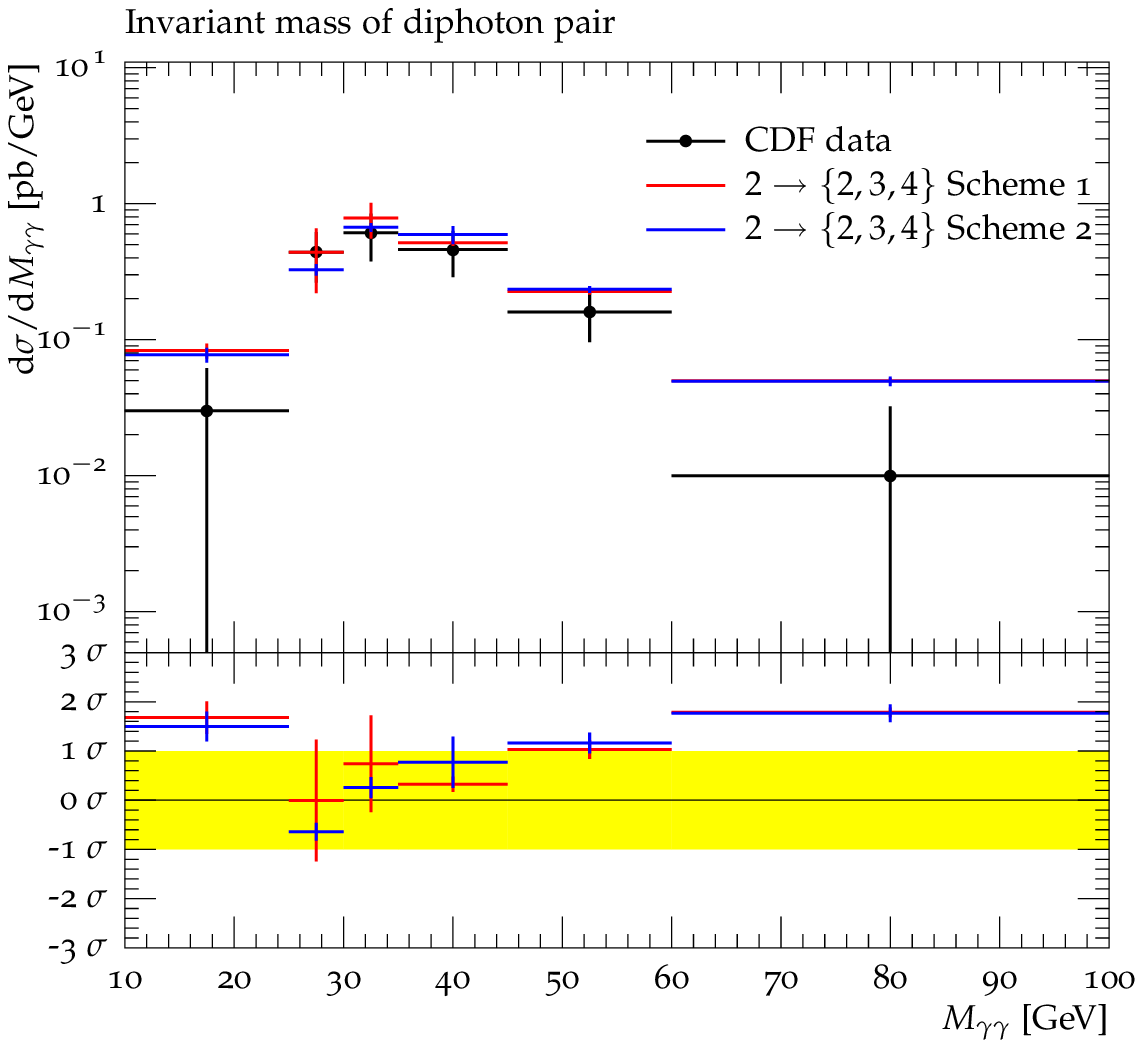}\\
    \includegraphics[width=\linewidth]{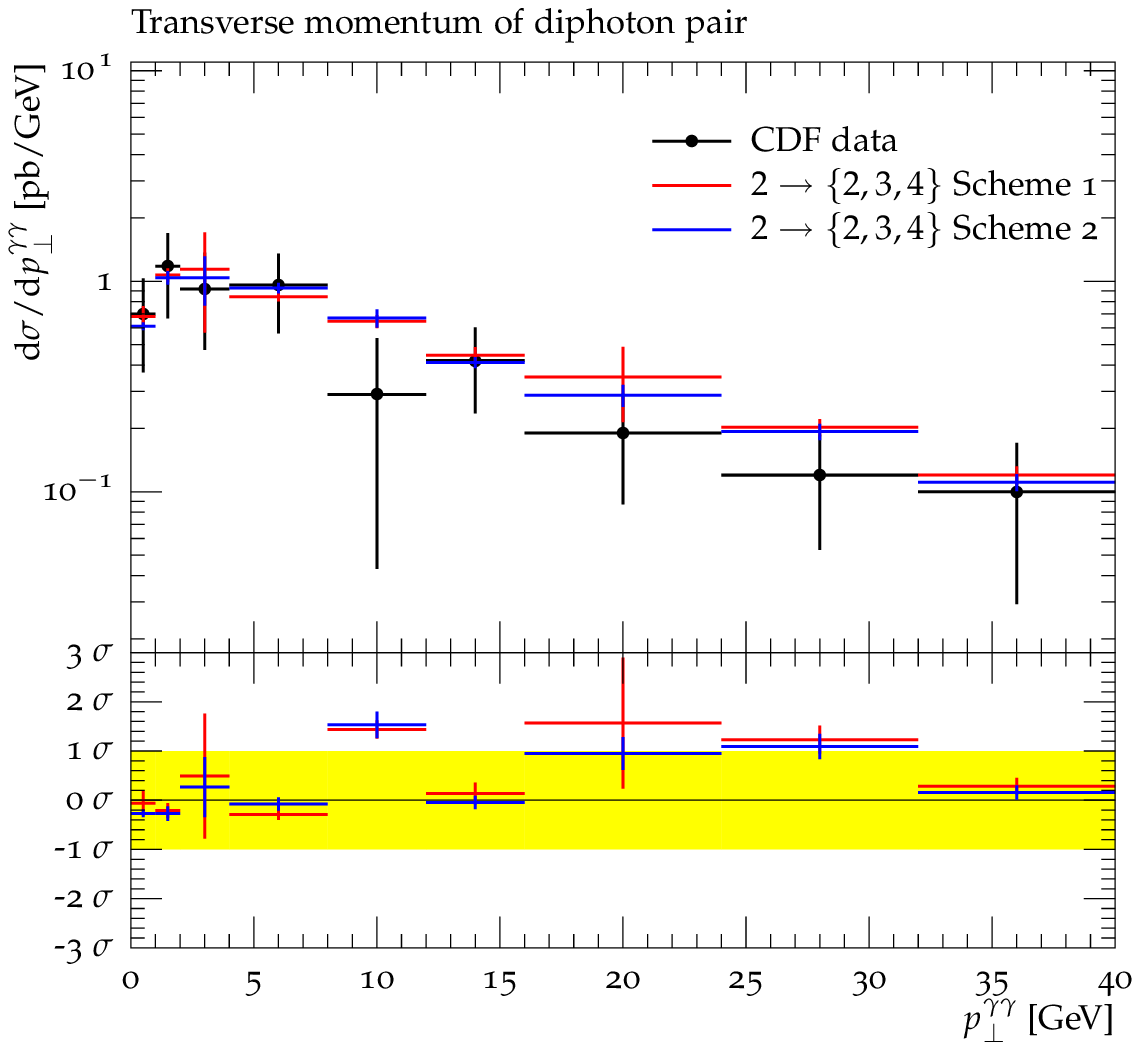}\\
    \includegraphics[width=\linewidth]{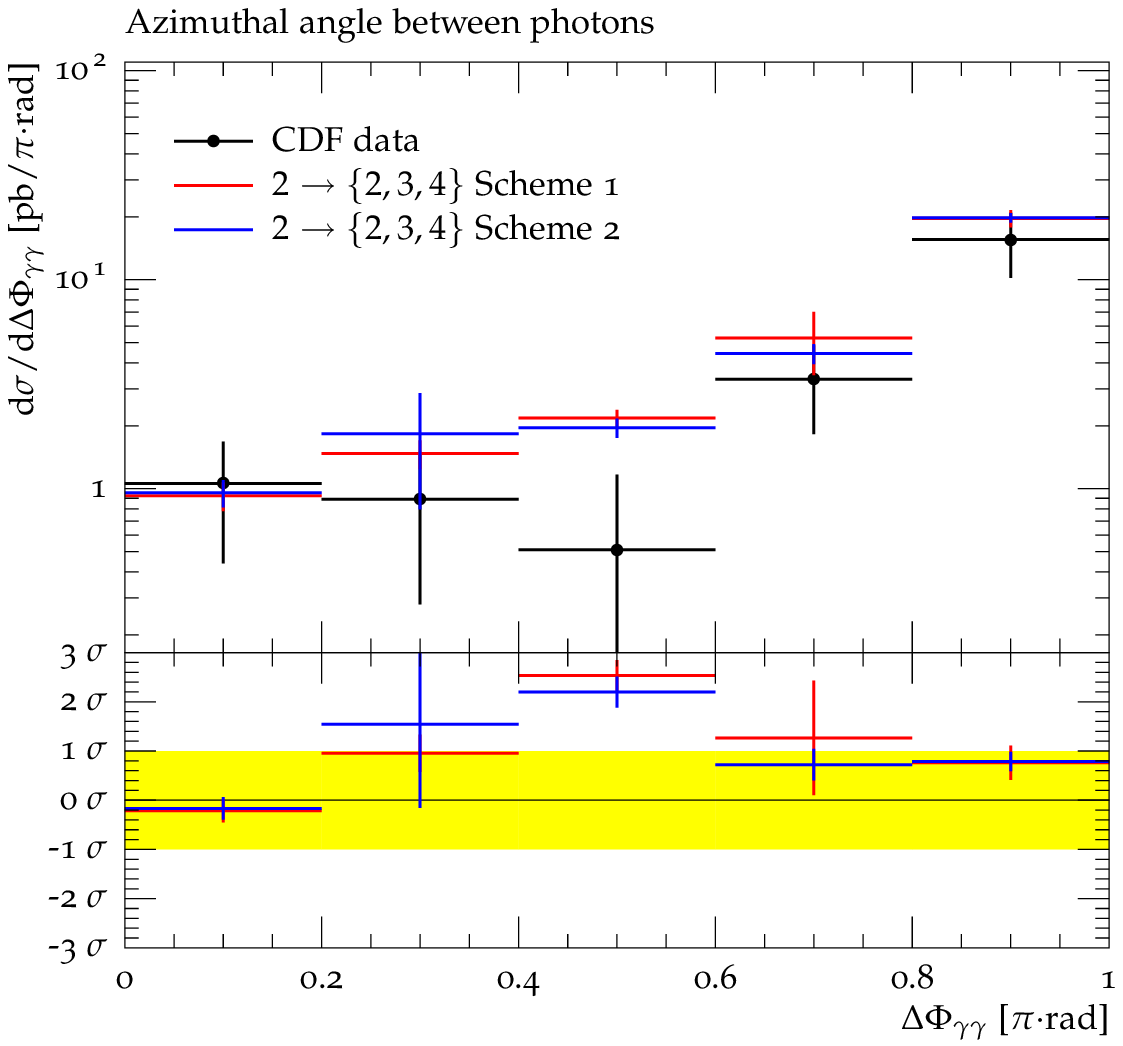}}
    \label{fig:tev_diphoton_me}}\hspace*{5mm}
  \caption{Properties of diphoton events measured by the CDF collaboration~\protect\cite{Acosta:2004sn}.
    Figure~\protect\subref{fig:tev_diphoton_ps} compares the influence of different
    parton-shower kinematics when using leading-order matrix elements.
    Figure~\protect\subref{fig:tev_diphoton_me} shows the same comparison
    for merged event samples with up to two additional particles in the matrix element-final state.
    Scheme 1 refers to the algorithm outlined in Appendix~\ref{sec:kinematics},
    scheme 2 stands for the original implementation~\protect\cite{Schumann:2007mg}.}
  \label{fig:tev_diphoton}
\end{centering}
\end{figure}

In this context the parton-shower kinematics might become
important, because the recoil scheme discussed in Sec.~\ref{sec:shower} plays an
important role for the generation of transverse momentum for the diphoton system.
Thus, as a first step, Figure~\ref{fig:tev_diphoton_ps} compares parton-level
Monte-Carlo predictions using two different splitting kinematics. We observe
that both, the algorithm outlined in Appendix~\ref{sec:kinematics}, denoted ``Scheme 1'', and the
method proposed in~\cite{Schumann:2007mg}, denoted ``Scheme 2'', have difficulties describing the
critical regions mentioned above.

We show in Fig.~\ref{fig:tev_diphoton_me} that with the inclusion of
higher-order real-emission matrix elements, the simulation is able to describe the
measurement significantly better. Especially the transverse-momentum
distribution exemplifies two unique features: The resummation of large
logarithms correctly reproduces the Sudakov-shape of the low-$p_\perp$ region
which is not possible with fixed-order calculations.
At the same time exact matrix elements allow for a consistent simulation
of the high-$p_\perp$ tail where a traditional parton-shower approach would fail.
Also the decorrelation between the photons is
now matched very well\footnote{The deviations in the second and third bin
might be due to large statistical fluctuations in the measurement and seem
to be contradicted by an earlier albeit unfortunately unpublished \DO{}
measurement~\cite{Chen:1997gg}.}.
The simulation becomes largely independent of the precise
implementation of the parton-shower kinematics, an effect which is also
observed in~\cite{Carli:2009cg}.
This happens because in the relevant part of the 
phase space, hard matrix elements are employed to define the kinematics of the 
diphoton system. In this way the merging algorithm
of Sec.~\ref{sec:meps} can be used to eliminate theoretical uncertainties 
in the parton-shower model employed. At the same time we substantiate our introductory statement,
that within the domain of hard-photon production the parton shower can easily be 
corrected using higher-order matrix elements.

\section{Conclusions}
\label{sec:conclusions}
We have presented Monte-Carlo algorithms for the precise simulation of hard-photon production
in collider experiments. Using interleaved QCD+QED evolution in a dipole-like parton shower 
enables us to simulate the photon fragmentation function within a general-purpose Monte-Carlo
event generator. Comparison with data from the ALEPH experiment exemplifies the quality of the 
approach.

To diminish intrinsic uncertainties of parton-shower models, we have supplemented our simulation 
with higher-order tree-level matrix elements. To do so, an existing algorithm for QCD 
has been extended to include QED democratically. This can be seen as a first step towards 
a unified prescription for treating strong and electroweak radiative corrections.
At the same time it provides a natural way to simulate hard-photon production, where 
the fragmentation component is described consistently by the combined QCD+QED resummation 
in the parton shower supplemented with a non-perturbative hadronisation model. We have 
employed this procedure to analyse prompt-photon production in hadron-hadron collisions and 
find an improved description of Tevatron measurements. Due to a much larger phase space available 
for radiative corrections, these effects should become even more pronounced at LHC energies.

\section*{Acknowledgements}
We would like to thank Thomas Binoth, Gudrun Heinrich, Thomas Gehrmann and Frank Krauss 
for numerous fruitful discussions.
Help from Thomas Binoth with the code for $gg \to \gamma\gamma$ is gratefully acknowledged.
We are grateful to Frank Krauss and Thomas Gehrmann for
valuable comments on the manuscript. 
We also thank Dmitry Bandurin and Michael Begel for their clarifications about
the \DO{} measurement. FS would like to thank the ITP Z\"urich for its hospitality
while this work was finalised.
SH acknowledges funding by the Swiss National Science Foundation (SNF, contract 
number 200020-126691) and by the University of Z{\"u}rich (Forschungskredit number 57183003). 
The work of FS was supported by the MCnet 
Marie Curie Research Training Network (contract number MRTN-CT-2006-035606).
SS acknowledges financial support by BMBF. 
\appendix
\section{Dipole-splitting kinematics}
\label{sec:kinematics}
\myfigure{t}{
  \scalebox{0.9}{\begin{picture}(180,90)(20,10)
    \Line(168, 72)(100, 50)
    \Line( 32, 72)(100, 50)
    \LongArrow( 75,51)( 50,59)
    \LongArrow(125,51)(150,59)
    \Photon(100,15)(100,40){3}{3}
    \GCirc(100,50){10}{1}
    \put(137, 43){$p_{\widetilde{ij}}$}
    \put( 20, 75){$\tilde{k}$}
    \put(175, 75){$\widetilde{ij}$}
    \put( 97,  3){$Q$}
    \put( 56, 43){$p_{\tilde{k}}$}
  \end{picture}}
  \begin{pspicture}(1,1)
    \psline[linewidth=0.05]{->}(0,0.5)(1,0.5)
  \end{pspicture}
  \scalebox{0.9}{\begin{picture}(200,90)(0,10)
    \Line(190, 80)(100, 50)
    \Line( 32, 72)(100, 50)
    \LongArrow( 75,51)( 50,59)
    \LongArrow(118,48)(138,55)
    \LongArrow(138,74)(112,83)
    \LongArrow(158,62)(184, 71)
    \Line(145, 65)(100, 80)
    \Vertex(145,65){2.5}
    \Photon(100,15)(100,40){3}{3}
    \GCirc(100,50){10}{1}
    \put(130, 40){$p_{ij}$}
    \put( 20, 75){$k$}
    \put( 90, 80){$j$}
    \put(197, 81){$i$}
    \put( 97,  3){$Q$}
    \put(172, 55){$p_i$}
    \put( 56, 43){$p_k$}
    \put(130, 87){$p_j$}
  \end{picture}}
}{Schematic view of the splitting of a final-state parton with a 
  final-state spectator.  The blob denotes the hard process from which the
  three partons emerge. Since the four-momentum of the splitter-spectator 
  system is conserved, the pair can be regarded as decay products in a
  two-particle transition of a virtual particle with momentum $Q=p_{ij}+p_k$.
  \label{fig:split_ff}}

In this appendix we derive alternative splitting kinematics for dipole-like 
parton showers in the spirit of~\cite{Schumann:2007mg}. In addition to the massless case
proposed in~\cite{Platzer:2009jq} we give explicit formulae for the fully massive case,
which plays an important role for truncated-shower algorithms and in processes involving 
heavy quarks~\cite{Aivazis:1993pi}.
Effectively, only one dipole configuration is considered, i.e.\ branching 
final-state partons with the spectator parton being in the final state.
In this case we closely follow the proposal in~\cite{Schumann:2007mg}, 
which is inspired by the original kinematics of the massive Catani-Seymour 
dipoles~\cite{Catani:1996vz,*Catani:2002hc}. Kinematic relations for all other 
dipole configurations are then derived using crossing relations.

We consider the process depicted in Fig.~\ref{fig:split_ff}, where a parton 
$\widetilde{ij}$, accompanied by a spectator parton $\tilde{k}$, splits into partons 
$i$ and $j$, with the recoil absorbed by the spectator $k$.
We define the combined momenta $p_{ij}=p_i+p_j$ and $Q=p_{ij}+p_k$ and the variables
\begin{align}
  \eta_{ij,k}\,=&\;\frac{p_ip_j}{p_{ij}p_k}\;,
  &\xi_{i,jk}\,=&\;\frac{p_ip_k}{p_{ij}p_k}\;.
\end{align}
Thus, we immediately obtain
\begin{equation}\label{eq:def_sij}
  s_{ij}\,=\;\frac{\eta_{ij,k}}{1+\eta_{ij,k}}\,(Q^2-m_k^2)
    +\frac{1}{1+\eta_{ij,k}}\,(m_i^2+m_j^2)\;.\\
\end{equation}
Now let the light-like helper vectors $l$ and $n$ be given by
(cf.~\cite{Pittau:1996ez,*Pittau:1997mv})
\begin{align}\label{eq:sudakov_momenta}
  l\,=&\;\frac{p_{ij}-\alpha_{ij}\, p_k}{1-\alpha_{ij}\alpha_k}\,\;,
  &n\,=&\;\frac{p_k-\alpha_k\, p_{ij}}{1-\alpha_{ij}\alpha_k}\,\;,
\end{align}
where
\begin{align}
  \alpha\,=&\;\frac{p^2}{\gamma_{ij,k}}\;
  &&\text{and}
  &\gamma_{ij,k}\,=&\;2\,ln\,=\;\frac{1}{2}\sbr{\,\rbr{Q^2-s_{ij}-m_k^2}
    +{\rm sgn}\rbr{Q^2-s_{ij}-m_k^2}\sqrt{\lambda(Q^2,s_{ij},m_k^2)}\;}\;,
\end{align}
with $\lambda$ denoting the K{\"a}llen function $\lambda(a,b,c)=\rbr{a-b-c}^2-4\,bc$.\\
The momenta $p_i$ and $p_j$ can then be expressed in terms of $l$, 
$n$ and a transverse component $k_\perp$ as
\begin{align}\label{eq:def_pi_pj_FF}
  p_i^\mu\,=&\;z_i\,l^\mu+\frac{m_i^2+{\rm k}_\perp^2}{z_i}\,
    \frac{n^\mu}{2\,ln}+k_\perp^\mu\;,
  &p_j^\mu\,=&\;(1-z_i)\,l^\mu+\frac{m_j^2+{\rm k}_\perp^2}{1-z_i}\,
    \frac{n^\mu}{2\,ln}-k_\perp^\mu\;,
\end{align}
The parameters $z_i$ and ${\rm k}_\perp^2$ of this decomposition are given by
\begin{equation}\label{eq:def_zi_kt}
  \begin{split}
    z_i\,=&\;\frac{Q^2-s_{ij}-m_k^2}{
      \sqrt{\lambda\rbr{Q^2,s_{ij},m_k^2}}}\,
      \sbr{\;\xi_{i,jk}\,-\,\frac{m_k^2}{\abs{\gamma_{ij,k}}}\rbr{\,
        \eta_{ij,k}+\frac{2\,m_i^2}{Q^2-s_{ij}-m_k^2}}\,}\;,\\
    {\rm k}_\perp^2\,=&\;\rbr{Q^2-m_i^2-m_j^2-m_k^2}\,
      \frac{\eta_{ij,k}}{1+\eta_{ij,k}}\;z_i\,(1-z_i)
      -(1-z_i)^2 m_i^2-z_i^2 m_j^2\;.
  \end{split}
\end{equation}
Equations~\eqref{eq:def_sij} and~\eqref{eq:def_zi_kt} are valid for all dipole
configurations, i.e.\ initial and final-state branchings with the recoil partner being 
either in the initial or in the final state. The corresponding mapping of variables
$\eta_{ij,k}$ and $\xi_{i,jk}$ onto those defined for massless partons
in~\cite{Catani:1996vz,*Catani:2002hc} is listed in Tab.~\ref{tab:def_eta_xi}.
Note, that in the massless case these kinematic relations reproduce the results
of~\cite{Platzer:2009jq}.

\mytable{t}{
  \begin{tabular}{|c|c|c|c|c|c|c|}
  \cline{1-3}\cline{5-7}
  \vphantom{$\dst\int$}Configuration &
    \hspace*{9mm}$\xi_{i,jk}$\hspace*{9mm} & 
    \hspace*{7mm}$\eta_{ij,k}$\hspace*{7mm} & &
    Configuration & 
    \hspace*{9mm}$\xi_{j,ak}$\hspace*{9mm} & 
    \hspace*{7mm}$\eta_{ja,k}$\hspace*{7mm} \\
  \cline{1-3}\cline{5-7}
  \vphantom{$\dst\int\limits_A^B$}FF & 
    $\tilde{z}_i$ & $\dst\frac{y_{ij,k}}{1-y_{ij,k}}$ & &
  \vphantom{$\dst\int\limits_A^B$}IF & 
    $\abs{\,1-\dst\frac{1-u_j}{x_{jk,a}-u_j}\,}$ &
    $\dst\frac{u_j}{x_{jk,a}-u_j}$ \\
  \cline{1-3}\cline{5-7}
  \vphantom{$\dst\int\limits_A^B$}FI & 
    $-\,\tilde{z}_i$ & $x_{ij,a}-1$ & &
  \vphantom{$\dst\int\limits_A^B$}II & 
    $\dst1-\frac{1}{x_{j,ab}+\tilde{v}_j}$ &
    $\dst\frac{-\,\tilde{v}_j}{x_{j,ab}+\tilde{v}_j}$ \\
  \cline{1-3}\cline{5-7}
  \end{tabular}
  }{Mapping of variables for Eqs.~\eqref{eq:def_sij} and~\eqref{eq:def_zi_kt}.
  \label{tab:def_eta_xi}
}

Constructing a shower emission proceeds as follows
\begin{enumerate}
\item 
Determine $\eta_{ij,k}$ and $\xi_{i,jk}$ from evolution and splitting variable.\\
Compute $s_{ij}$ according to Eq.~\eqref{eq:def_sij} and $z_i$ and ${\rm k}_\perp^2$ 
according to Eqs.~\eqref{eq:def_zi_kt}.
\item Construct $p_k$ according to\footnote{Relation~\eqref{eq:def_pk} is crossing invariant 
  and can therefore be employed for all dipole configurations.}~\cite{Catani:1996vz,*Catani:2002hc}
\begin{equation}\label{eq:def_pk}
  \begin{split}
  p_k=&\;\left(\tilde{p}_k-\frac{Q^2+m_k^2-m_{ij}^2}{2\,Q^2}\,Q\right)\,
    \sqrt{\frac{\lambda(Q^2,s_{ij},m_k^2)}{\lambda(Q^2,m_{ij}^2,m_k^2)}}
    +\frac{Q^2+m_k^2-s_{ij}}{2\,Q^2}\,Q\;.
  \end{split}
\end{equation}
\item From $p_{ij}$ and $p_k$ construct the light-like momenta $l$ and $n$ 
according to Eqs.~\eqref{eq:sudakov_momenta}.\\
Determine the new momenta $p_i$ and $p_j$ according to Eqs.~\eqref{eq:def_pi_pj_FF}.
\end{enumerate}

Due to the possibly vanishing denominator of $\xi_{i,jk}$ 
and $\eta_{ij,k}$ in the case of initial-state splitters with final-state spectator,
the corresponding kinematic relations occasionally become numerically unstable.
Except for a tiny region of the phase space, where $\abs{x_{ik,a}-u_i}<\epsilon$,
such configurations can be dealt with in the following way:
Instead of constructing $p_i$, we construct $(x_{ik,a}-u_i)\,p_i\;$. The 
energy of this helper momentum is determined using the on-shell constraint 
for $p_i$ and the whole four-vector is then rescaled with $1/(x_{ik,a}-u_i)$.
We typically have $\epsilon$ of the order of $10^{-3}$.

\section{Enhancing photon production in the parton shower}
\label{sec:enhance}

To improve the statistical significance of event samples with identified photons
described by the parton shower, it is useful to enhance the corresponding 
branching probabilities. When doing so, one must of course correct for this enhancement
by means of an event weight, which depends on both, acceptance and rejection probabilities
in the parton-shower evolution. In this appendix we describe a method to incorporate
an enhancement, which is constant over the emission phase space. Our derivation is based 
on the applicability of the veto algorithm, cf.\ e.g.~\cite{Sjostrand:2006za}.

Let $t$ be the parton-shower evolution variable and $f(t)$ the splitting kernel
$\mc{K}$, integrated over the splitting variable $z$.\footnote{For simplicity,
  we assume that only one splitting function exists, i.e.\ that there is no flavour change 
  of the splitter during the evolution. The extension to flavour changing splittings 
  is straightforward, but it would unnecessarily complicate the notation at this point.}
The differential probability for generating a branching at scale $t$, when starting
from an upper evolution scale $t'$ is then given by
\begin{equation}\label{eq:va_prob}
  \mc{P}(t,t')\,=\;f(t)\,\exp\cbr{-\int_t^{t'}\done\bar{t}\,f(\bar{t})}\;.
\end{equation}
A new scale $t$ is therefore found as
\begin{align}
  t\,=&\;F^{-1}\sbr{\,F(t')+\log R\,}
  &&\text{where}
  &F(t)\,=\,\int^t\done t\,f(t)\;,
\end{align}
and where $R$ is a random number between zero and one.
The key point of the veto algorithm is, that even if the integral $F(t)$ is unknown, one
can still generate events according to $\mc{P}$ using an overestimate $g(t)\ge f(t)$
with a known integral $G(t)$. Firstly, a value $t$ is generated as $t=G^{-1}\sbr{\,G(t')+\log R\,}$.
Secondly, the value is accepted with probaility $f(t)/g(t)$. 
A splitting at $t$ with $n$ intermediate rejections is then produced with differential 
probability
\begin{equation}\label{eq:va_mcprob}
  \mc{P}_n(t,t')\,=\;\frac{f(t)}{g(t)}\,g(t)\,\exp\cbr{-\int_t^{t_1}\done\bar{t}\,g(\bar{t})}
    \prod_{i=1}^n\sbr{\,\int_{t_{i-1}}^{t_{i+1}}\done t_i\rbr{1-\frac{f(t_i)}{g(t_i)}}g(t_i)\,
      \exp\cbr{-\int_{t_i}^{t_{i+1}}\done\bar{t}\,g(\bar{t})}}\;,
\end{equation}
where $t_{n+1}=t'$ and $t_0=t$. The nested integrals in Eq.~\eqref{eq:va_mcprob} can be disentangled, and summing 
over $n$ leads to the exponentiation of the factor $g(t)-f(t)$, such that Eq.~\eqref{eq:va_prob}
is reproduced~\cite{Sjostrand:2006za}.

Our purpose is to introduce an additional overestimate $h(t)=C\,g(t)$, where $C$ is a constant.
The additional weight $g(t)/h(t)=1/C$ is then applied analytically rather than using a hit-or-miss method.
This leads to the following expression for the differential
probability to generate an emission at $t$ with $n$ rejections between $t$ and $t'$
\begin{equation}\label{eq:wva_mcprob}
  \begin{split}
  \mc{P}_n(t,t')\,=&\;\frac{f(t)}{g(t)}\,h(t)\,\exp\cbr{-\int_t^{t_1}\done\bar{t}\,h(\bar{t})}
    \prod_{i=1}^n\sbr{\,\int_{t_{i-1}}^{t_{i+1}}\done t_i\rbr{1-\frac{f(t_i)}{g(t_i)}}h(t_i)\,
      \exp\cbr{-\int_{t_i}^{t_{i+1}}\done\bar{t}\,h(\bar{t})}}\\
  &\quad\times\;\frac{1}{C}\,\prod_{i=1}^n\frac{g(t_i)-f(t_i)/C}{g(t_i)-f(t_i)}\;.
  \end{split}
\end{equation} 
The factor in the second line of Eq.~\eqref{eq:wva_mcprob} gives the analytic weight
associated with this event, where the term $1/C$ is due to the acceptance of the emission
with probability $f(t)/h(t)$. The product, which is needed for an exponentiation of $h(t)-f(t)$ 
instead of $g(t)-f(t)$, runs over all correction weights for rejected steps.

\section{Monte-Carlo setup}
\label{sec:sherpa}

\newcommand{\resummed}{r}

In this appendix we describe the details of the Monte-Carlo setup used to
generate the results in this publication. We employ the
\Sherpa\cite{Gleisberg:2003xi,*Gleisberg:2008ta} framework, which is a multi-purpose
Monte-Carlo event generator for collider experiments.

The matrix-element generator \Comix\cite{Gleisberg:2008fv} is used to
produce parton-level events for the following processes:\\[3mm]
\begin{tabular}{{l}{r}{c}{l}{l}}
  $e^+ e^-$ collisions (Sec.~\ref{sec:results:ee}) & 
  $e^+ e^-$ & $\to$ & $\resummed \resummed + N \resummed{}$ & $N\leq N_\mathrm{max}$ \\
  $p \bar{p}$ collisions (Sec.~\ref{sec:results:tev1}, \ref{sec:results:tev2}) & 
  $p \bar{p}$ & $\to$ & $\resummed \resummed + N \resummed{}$ & $N\leq N_\mathrm{max}$ \\
\end{tabular}\\[3mm]
The ``resummed'' container $\resummed$ implements the democratic treatment
of photon and parton radiation, i.e. it contains the light quarks $d$, $u$, $s$, $c$ 
and $b$ as well as gluons and photons.
In addition to all automatically generated tree-level amplitudes,
the loop induced process $g g \to \gamma \gamma$\cite{Berger:1983yi} has been
implemented.

As a parton shower we employ the \CSS~\cite{Schumann:2007mg} module which in case
of $N_\mathrm{max}>0$ is merged to the matrix-element emissions above $\qcut=10$ GeV through the algorithm
described in~\cite{Hoeche:2009rj}. Unless stated otherwise, we employ the parton shower
kinematics described in Appendix~\ref{sec:kinematics}. All QED splitting functions
are included. The shower cut-off has been left at the default value of
$p_\perp^{\rm min}=1$ GeV for the LEP runs and has been switched to
$p_\perp^{\rm min}=2$ GeV for the Tevatron runs purely for efficiency reasons.

The PDF set employed for $p \bar{p}$ runs is CTEQ~6L~\cite{Pumplin:2002vw}, which
defines the corresponding $\alpha_s$ parametrisation in hadron collisions.
All other generator parameters are left at the default values of the Monte Carlo
programs, since none of them is expected to have any impact on the results presented here.

Hadron-level results for the fragmentation function analysis in Sec.~\ref{sec:results:ee} are
produced using the fragmentation module \Ahadic~\cite{Winter:2003tt,*Krauss:2010xy} and the
hadron and $\tau$ decay package \Hadrons~\cite{Krauss:2010xx}. 
The \Ahadic{} default tune to data from the LEP experiments at the $Z^0$ resonance, obtained
using the \Professor{}~\cite{Buckley:2009bj} framework, has been used.
To account for corrections in
the ALEPH measurement, the decays of $\pi^0$ and $\eta$ have been disabled.
Extra QED radiation in hadron decays is simulated through \Photons~\cite{Schonherr:2008av}.
All Tevatron analyses are presented at the parton level after parton-shower evolution.

Multiple parton interactions have been disabled, because the presented
measurements have been corrected for their effects.

\bibliographystyle{bib/amsunsrt_mod}  
\bibliography{bib/journal}

\begin{thebibliography}{10}

\bibitem{Abazov:2008it}
V.~M. Abazov et~al., D0 collaboration, \emph{{Search for Decay of a
  Fermiophobic Higgs Boson $h_f \to \gamma \gamma$ with the D0 Detector at
  $\sqrt{s}$ = 1.96 TeV}}, Phys. Rev. Lett. \textbf{101} (2008),
  \href{http://www.slac.stanford.edu/spires/find/hep/www?eprint=0803.1514}{051%
801},  [\href{http://arXiv.org/pdf/0803.1514}{{\tt arXiv:0803.1514}} [hep-ex]].
  \relax
 \relax
\bibitem{Aaltonen:2009ga}
T.~Aaltonen et~al., CDF collaboration, \emph{{Search for a Fermiophobic Higgs
  Boson Decaying into Diphotons in $p\bar{p}$ Collisions at $\sqrt{s} = 1.96$
  TeV}}, Phys. Rev. Lett. \textbf{103} (2009),
  \href{http://www.slac.stanford.edu/spires/find/hep/www?eprint=0905.0413}{061%
803},  [\href{http://arXiv.org/pdf/0905.0413}{{\tt arXiv:0905.0413}} [hep-ex]].
  \relax
 \relax
\bibitem{Aad:2009wy}
\href{http://www-spires.dur.ac.uk/spires/find/hep/www?eprint=arXiv:0901.0512}{%
G.~Aad et~al.}, The ATLAS collaboration, \emph{{Expected Performance of the
  ATLAS Experiment - Detector, Trigger and Physics}},
  \href{http://arXiv.org/pdf/0901.0512}{{\tt arXiv:0901.0512}} [hep-ex]. \relax
 \relax
\bibitem{Ball:2007zza}
G.~L. Bayatian et~al., CMS collaboration, \emph{{CMS technical design report,
  volume II: Physics performance}}, J. Phys. \textbf{G34} (2007),
  \href{http://www.slac.stanford.edu/spires/find/hep/www?j=J%20Phys,G34,995}{9%
95--1579}. \relax
 \relax
\bibitem{Rainwater:1997dg}
D.~L. Rainwater and D.~Zeppenfeld, \emph{{Searching for $H \to \gamma \gamma$
  in weak boson fusion at the LHC}}, JHEP \textbf{12} (1997),
  \href{http://www-spires.dur.ac.uk/spires/find/hep/www?eprint=hep-ph/9712271}%
{005},  [\href{http://arXiv.org/pdf/hep-ph/9712271}{{\tt hep-ph/9712271}}].
  \relax
 \relax
\bibitem{Abdullin:1998er}
S.~Abdullin et~al., \emph{{Higgs boson discovery potential of LHC in the
  channel $p p \to \gamma \gamma + jet$}}, Phys. Lett. \textbf{B431} (1998),
  \href{http://www.slac.stanford.edu/spires/find/hep/www?eprint=hep-ph/9805341%
}{410--419},  [\href{http://arXiv.org/pdf/hep-ph/9805341}{{\tt
  hep-ph/9805341}}]. \relax
 \relax
\bibitem{Hinchliffe:1998ys}
I.~Hinchliffe and F.~E. Paige, \emph{{Measurements in gauge mediated SUSY
  breaking models at LHC}}, Phys. Rev. \textbf{D60} (1999),
  \href{http://www.slac.stanford.edu/spires/find/hep/www?eprint=hep-ph/9812233%
}{095002},  [\href{http://arXiv.org/pdf/hep-ph/9812233}{{\tt hep-ph/9812233}}].
  \relax
 \relax
\bibitem{Giudice:1998ck}
G.~F. Giudice, R.~Rattazzi and J.~D. Wells, \emph{{Quantum gravity and extra
  dimensions at high-energy colliders}}, Nucl. Phys. \textbf{B544} (1999),
  \href{http://www.slac.stanford.edu/spires/find/hep/www?eprint=hep-ph/9811291%
}{3--38},  [\href{http://arXiv.org/pdf/hep-ph/9811291}{{\tt hep-ph/9811291}}].
  \relax
 \relax
\bibitem{Davoudiasl:2000wi}
H.~Davoudiasl, J.~L. Hewett and T.~G. Rizzo, \emph{{Experimental probes of
  localized gravity: On and off the wall}}, Phys. Rev. \textbf{D63} (2001),
  \href{http://www.slac.stanford.edu/spires/find/hep/www?eprint=hep-ph/0006041%
}{075004},  [\href{http://arXiv.org/pdf/hep-ph/0006041}{{\tt hep-ph/0006041}}].
  \relax
 \relax
\bibitem{Macesanu:2002ew}
C.~Macesanu, C.~D. McMullen and S.~Nandi, \emph{{New signal for universal extra
  dimensions}}, Phys. Lett. \textbf{B546} (2002),
  \href{http://www.slac.stanford.edu/spires/find/hep/www?eprint=hep-ph/0207269%
}{253--260},  [\href{http://arXiv.org/pdf/hep-ph/0207269}{{\tt
  hep-ph/0207269}}]. \relax
 \relax
\bibitem{Bhatti:2005ai}
A.~Bhatti et~al., \emph{{Determination of the jet energy scale at the Collider
  Detector at Fermilab}}, Nucl. Instrum. Meth. \textbf{A566} (2006),
  \href{http://www.slac.stanford.edu/spires/find/hep/www?eprint=hep-ex/0510047%
}{375--412},  [\href{http://arXiv.org/pdf/hep-ex/0510047}{{\tt
  hep-ex/0510047}}]. \relax
 \relax
\bibitem{Abbott:1998xw}
B.~Abbott et~al., D0 collaboration, \emph{{Determination of the absolute jet
  energy scale in the D\O\ calorimeters}}, Nucl. Instrum. Meth. \textbf{A424}
  (1999),
  \href{http://www.slac.stanford.edu/spires/find/hep/www?eprint=hep-ex/9805009%
}{352--394},  [\href{http://arXiv.org/pdf/hep-ex/9805009}{{\tt
  hep-ex/9805009}}]. \relax
 \relax
\bibitem{Golutvin:2008zz}
I.~A. Golutvin et~al., \emph{{Setting the Jet Energy Scale in the CMS
  Calorimeter using Events with Direct Photons}}, Phys. Part. Nucl. Lett.
  \textbf{5} (2008),
  \href{http://www.slac.stanford.edu/spires/find/hep/www?j=Phys%20Part%20Nucl%%
20Lett,5,447}{447--455}. \relax
 \relax
\bibitem{Aurenche:1988vi}
P.~Aurenche, R.~Baier, M.~Fontannaz, J.~F. Owens and M.~Werlen, \emph{{Gluon
  content of the nucleon probed with real and virtual photons}}, Phys. Rev.
  \textbf{D39} (1989),
  \href{http://www-spires.dur.ac.uk/spires/find/hep/www?j=PHRVA,D39,3275}{3275%
}. \relax
 \relax
\bibitem{Vogelsang:1995bg}
W.~Vogelsang and A.~Vogt, \emph{{Constraints on the proton's gluon distribution
  from prompt photon production}}, Nucl. Phys. \textbf{B453} (1995),
  \href{http://www-spires.dur.ac.uk/spires/find/hep/www?eprint=hep-ph/9505404}%
{334--354},  [\href{http://arXiv.org/pdf/hep-ph/9505404}{{\tt
  hep-ph/9505404}}]. \relax
 \relax
\bibitem{Fontannaz:2003yn}
M.~Fontannaz and G.~Heinrich, \emph{{Isolated photon plus jet photoproduction
  as a tool to constrain the gluon distribution in the proton and the photon}},
  Eur. Phys. J. \textbf{C34} (2004),
  \href{http://www.slac.stanford.edu/spires/find/hep/www?eprint=hep-ph/0312009%
}{191--199},  [\href{http://arXiv.org/pdf/hep-ph/0312009}{{\tt
  hep-ph/0312009}}]. \relax
 \relax
\bibitem{LlewellynSmith:1978dc}
C.~H. Llewellyn~Smith, \emph{{QCD predictions for processes involving real
  photons}}, Phys. Lett. \textbf{B79} (1978),
  \href{http://www-spires.dur.ac.uk/spires/find/hep/www?j=PHLTA,B79,83}{83}.
  \relax
 \relax
\bibitem{Baer:1990ra}
H.~Baer, J.~Ohnemus and J.~F. Owens, \emph{{Next-to-leading-logarithm
  calculation of direct photon production}}, Phys. Rev. \textbf{D42} (1990),
  \href{http://www-spires.dur.ac.uk/spires/find/hep/www?j=PHRVA,D42,61}{61--71%
}. \relax
 \relax
\bibitem{Aurenche:1989gv}
P.~Aurenche, R.~Baier and M.~Fontannaz, \emph{{Prompt photon production at
  colliders}}, Phys. Rev. \textbf{D42} (1990),
  \href{http://www-spires.dur.ac.uk/spires/find/hep/www?j=PHRVA,D42,1440}{1440%
--1449}. \relax
 \relax
\bibitem{Glover:1993xc}
E.~W.~N. Glover and A.~G. Morgan, \emph{{Measuring the photon fragmentation
  function at LEP}}, Z. Phys. \textbf{C62} (1994),
  \href{http://www-spires.dur.ac.uk/spires/find/hep/www?j=ZEPYA,C62,311}{311--%
322}. \relax
 \relax
\bibitem{GehrmannDeRidder:1997wx}
A.~Gehrmann-De~Ridder, T.~Gehrmann and E.~W.~N. Glover, \emph{{Radiative
  corrections to the photon + 1 jet rate at LEP}}, Phys. Lett. \textbf{B414}
  (1997),
  \href{http://www-spires.dur.ac.uk/spires/find/hep/www?eprint=hep-ph/9705305}%
{354--361},  [\href{http://arXiv.org/pdf/hep-ph/9705305}{{\tt
  hep-ph/9705305}}]. \relax
 \relax
\bibitem{GehrmannDeRidder:1998ba}
A.~Gehrmann-De~Ridder and E.~W.~N. Glover, \emph{{Final state photon production
  at LEP}}, Eur. Phys. J. \textbf{C7} (1999),
  \href{http://www-library.desy.de/spires/find/hep/www?eprint=hep-ph/9806316}{%
29--48},  [\href{http://arXiv.org/pdf/hep-ph/9806316}{{\tt hep-ph/9806316}}].
  \relax
 \relax
\bibitem{Frixione:1998jh}
S.~Frixione, \emph{{Isolated photons in perturbative QCD}}, Phys. Lett.
  \textbf{B429} (1998),
  \href{http://www-spires.dur.ac.uk/spires/find/hep/www?eprint=hep-ph/9801442}%
{369--374},  [\href{http://arXiv.org/pdf/hep-ph/9801442}{{\tt
  hep-ph/9801442}}]. \relax
 \relax
\bibitem{Frixione:1999gr}
S.~Frixione and W.~Vogelsang, \emph{{Isolated-photon production in polarized $p
  p$ collisions}}, Nucl. Phys. \textbf{B568} (2000),
  \href{http://www-spires.dur.ac.uk/spires/find/hep/www?eprint=hep-ph/9908387}%
{60--92},  [\href{http://arXiv.org/pdf/hep-ph/9908387}{{\tt hep-ph/9908387}}].
  \relax
 \relax
\bibitem{Aurenche:1987fs}
P.~Aurenche, R.~Baier, M.~Fontannaz and D.~Schiff, \emph{{Prompt photon
  production at large $p_T$ Scheme invariant QCD predictions and comparison
  with experiment}}, Nucl. Phys. \textbf{B297} (1988),
  \href{http://www.slac.stanford.edu/spires/find/hep/www?j=Nucl%20Phys,B297,66%
1}{661}. \relax
 \relax
\bibitem{Gordon:1994ut}
L.~E. Gordon and W.~Vogelsang, \emph{{Polarized and unpolarized isolated prompt
  photon production beyond the leading order}}, Phys. Rev. \textbf{D50} (1994),
  \href{http://www.slac.stanford.edu/spires/find/hep/www?j=Phys%20Rev,D50,1901%
}{1901--1916}. \relax
 \relax
\bibitem{Aurenche:1985yk}
P.~Aurenche, A.~Douiri, R.~Baier, M.~Fontannaz and D.~Schiff, \emph{{Large
  $p_T$ Double Photon Production in Hadronic Collisions --- Beyond Leading
  Logarithm QCD Calculation}}, Z. Phys. \textbf{C29} (1985),
  \href{http://www.slac.stanford.edu/spires/find/hep/www?j=Z%20Phys,C29,459}{4%
59--475}. \relax
 \relax
\bibitem{Bailey:1992br}
B.~Bailey, J.~F. Owens and J.~Ohnemus, \emph{{Order $\alpha_s$ Monte Carlo
  calculation of hadronic double-photon production}}, Phys. Rev. \textbf{D46}
  (1992),
  \href{http://www.slac.stanford.edu/spires/find/hep/www?j=Phys%20Rev,D46,2018%
}{2018--2027}. \relax
 \relax
\bibitem{Catani:2002ny}
S.~Catani, M.~Fontannaz, J.~P. Guillet and E.~Pilon, \emph{{Cross section of
  isolated prompt photons in hadron-hadron collisions}}, JHEP \textbf{05}
  (2002),
  \href{http://www-spires.dur.ac.uk/spires/find/hep/www?eprint=hep-ph/0204023}%
{028},  [\href{http://arXiv.org/pdf/hep-ph/0204023}{{\tt hep-ph/0204023}}].
  \relax
 \relax
\bibitem{Aurenche:2006vj}
P.~Aurenche, M.~Fontannaz, J.-P. Guillet, E.~Pilon and M.~Werlen, \emph{{Recent
  critical study of photon production in hadronic collisions}}, Phys. Rev.
  \textbf{D73} (2006),
  \href{http://www-spires.dur.ac.uk/spires/find/hep/www?eprint=hep-ph/0602133}%
{094007},  [\href{http://arXiv.org/pdf/hep-ph/0602133}{{\tt hep-ph/0602133}}].
  \relax
 \relax
\bibitem{Belghobsi:2009hx}
Z.~Belghobsi et~al., \emph{{Photon-jet correlations and constraints on
  fragmentation functions}}, Phys. Rev. \textbf{D79} (2009),
  \href{http://www.slac.stanford.edu/spires/find/hep/www?eprint=0903.4834}{114%
024},  [\href{http://arXiv.org/pdf/0903.4834}{{\tt arXiv:0903.4834}} [hep-ph]].
  \relax
 \relax
\bibitem{Binoth:1999qq}
T.~Binoth, J.~P. Guillet, E.~Pilon and M.~Werlen, \emph{{A full next-to-leading
  order study of direct photon pair production in hadronic collisions}}, Eur.
  Phys. J. \textbf{C16} (2000),
  \href{http://www-spires.dur.ac.uk/spires/find/hep/www?eprint=hep-ph/9911340}%
{311--330},  [\href{http://arXiv.org/pdf/hep-ph/9911340}{{\tt
  hep-ph/9911340}}]. \relax
 \relax
\bibitem{Binoth:2000zt}
T.~Binoth, J.~P. Guillet, E.~Pilon and M.~Werlen, \emph{{Beyond leading order
  effects in photon pair production at the Fermilab Tevatron}}, Phys. Rev.
  \textbf{D63} (2001),
  \href{http://www.slac.stanford.edu/spires/find/hep/www?eprint=hep-ph/0012191%
}{114016},  [\href{http://arXiv.org/pdf/hep-ph/0012191}{{\tt hep-ph/0012191}}].
  \relax
 \relax
\bibitem{DelDuca:2003uz}
V.~Del~Duca, F.~Maltoni, Z.~Nagy and Z.~Trocsanyi, \emph{{QCD radiative
  corrections to prompt diphoton production in association with a jet at hadron
  colliders}}, JHEP \textbf{04} (2003),
  \href{http://www-spires.dur.ac.uk/spires/find/hep/www?eprint=hep-ph/0303012}%
{059},  [\href{http://arXiv.org/pdf/hep-ph/0303012}{{\tt hep-ph/0303012}}].
  \relax
 \relax
\bibitem{Berger:1983yi}
E.~L. Berger, E.~Braaten and R.~D. Field, \emph{{Large-$p_\mathrm{T}$
  Production of Single and Double Photons in Proton-Proton and Pion-Proton
  Collisions}}, Nucl. Phys. \textbf{B239} (1984),
  \href{http://www.slac.stanford.edu/spires/find/hep/www?j=Nucl%20Phys,B239,52%
}{52}. \relax
 \relax
\bibitem{deFlorian:1999tp}
D.~de~Florian and Z.~Kunszt, \emph{{Two photons plus jet at LHC: The NNLO
  contribution from the $g g$ initiated process}}, Phys. Lett. \textbf{B460}
  (1999),
  \href{http://www-spires.dur.ac.uk/spires/find/hep/www?eprint=hep-ph/9905283}%
{184--188},  [\href{http://arXiv.org/pdf/hep-ph/9905283}{{\tt
  hep-ph/9905283}}]. \relax
 \relax
\bibitem{Balazs:1999yf}
C.~Balazs, P.~M. Nadolsky, C.~Schmidt and C.~P. Yuan, \emph{{Diphoton
  background to Higgs boson production at the LHC with soft gluon effects}},
  Phys. Lett. \textbf{B489} (2000),
  \href{http://www-spires.dur.ac.uk/spires/find/hep/www?eprint=hep-ph/9905551}%
{157--162},  [\href{http://arXiv.org/pdf/hep-ph/9905551}{{\tt
  hep-ph/9905551}}]. \relax
 \relax
\bibitem{Binoth:2003xk}
T.~Binoth, J.~P. Guillet and F.~Mahmoudi, \emph{{A compact representation of
  the $\gamma \gamma g g g \to 0$ amplitude}}, JHEP \textbf{02} (2004),
  \href{http://www-spires.dur.ac.uk/spires/find/hep/www?eprint=hep-ph/0312334}%
{057},  [\href{http://arXiv.org/pdf/hep-ph/0312334}{{\tt hep-ph/0312334}}].
  \relax
 \relax
\bibitem{Balazs:1997hv}
C.~Balazs, E.~L. Berger, S.~Mrenna and C.~P. Yuan, \emph{{Photon pair
  production with soft gluon resummation in hadronic interactions}}, Phys. Rev.
  \textbf{D57} (1998),
  \href{http://www-spires.dur.ac.uk/spires/find/hep/www?eprint=hep-ph/9712471}%
{6934--6947},  [\href{http://arXiv.org/pdf/hep-ph/9712471}{{\tt
  hep-ph/9712471}}]. \relax
 \relax
\bibitem{Balazs:2006cc}
C.~Balazs, E.~L. Berger, P.~M. Nadolsky and C.~P. Yuan, \emph{{All-orders
  resummation for diphoton production at hadron colliders}}, Phys. Lett.
  \textbf{B637} (2006),
  \href{http://www-spires.dur.ac.uk/spires/find/hep/www?eprint=hep-ph/0603037}%
{235--240},  [\href{http://arXiv.org/pdf/hep-ph/0603037}{{\tt
  hep-ph/0603037}}]. \relax
 \relax
\bibitem{Balazs:2007hr}
C.~Balazs, E.~L. Berger, P.~M. Nadolsky and C.~P. Yuan, \emph{{Calculation of
  prompt diphoton production cross sections at Fermilab Tevatron and CERN LHC
  energies}}, Phys. Rev. \textbf{D76} (2007),
  \href{http://www-spires.dur.ac.uk/spires/find/hep/www?eprint=arXiv:0704.0001%
}{013009},  [\href{http://arXiv.org/pdf/0704.0001}{{\tt arXiv:0704.0001}}
  [hep-ph]]. \relax
 \relax
\bibitem{Kidonakis:1999hq}
N.~Kidonakis and J.~F. Owens, \emph{{Soft-gluon resummation and NNLO
  corrections for direct photon production}}, Phys. Rev. \textbf{D61} (2000),
  \href{http://www-spires.dur.ac.uk/spires/find/hep/www?eprint=hep-ph/9912388}%
{094004},  [\href{http://arXiv.org/pdf/hep-ph/9912388}{{\tt hep-ph/9912388}}].
  \relax
 \relax
\bibitem{deFlorian:2005wf}
D.~de~Florian and W.~Vogelsang, \emph{{Threshold resummation for the
  prompt-photon cross section revisited}}, Phys. Rev. \textbf{D72} (2005),
  \href{http://www-spires.dur.ac.uk/spires/find/hep/www?eprint=hep-ph/0506150}%
{014014},  [\href{http://arXiv.org/pdf/hep-ph/0506150}{{\tt hep-ph/0506150}}].
  \relax
 \relax
\bibitem{Diana:2009xv}
G.~Diana, \emph{{High-energy resummation in direct photon production}}, Nucl.
  Phys. \textbf{B824} (2010),
  \href{http://www-spires.dur.ac.uk/spires/find/hep/www?eprint=arXiv:0906.4159%
}{154--167},  [\href{http://arXiv.org/pdf/0906.4159}{{\tt arXiv:0906.4159}}
  [hep-ph]]. \relax
 \relax
\bibitem{Becher:2009th}
T.~Becher and M.~D. Schwartz, \emph{{Direct photon production with effective
  field theory}}, JHEP \textbf{02} (2010),
  \href{http://www-spires.dur.ac.uk/spires/find/hep/www?eprint=arXiv:0911.0681%
}{040},  [\href{http://arXiv.org/pdf/0911.0681}{{\tt arXiv:0911.0681}}
  [hep-ph]]. \relax
 \relax
\bibitem{Yennie:1961ad}
D.~R. Yennie, S.~C. Frautschi and H.~Suura, \emph{{The Infrared Divergence
  Phenomena and High-Energy Processes}}, Ann. Phys. \textbf{13} (1961),
  \href{http://www.slac.stanford.edu/spires/find/hep/www?j=APNYA,13,379}{379--%
452}. \relax
 \relax
\bibitem{Summers:1994mc}
D.~J. Summers, \emph{{Exponentiation of soft photons in a process involving
  hard photons}}, Phys. Rev. \textbf{D53} (1996),
  \href{http://www-spires.dur.ac.uk/spires/find/hep/www?eprint=hep-ph/9405430}%
{2430--2441},  [\href{http://arXiv.org/pdf/hep-ph/9405430}{{\tt
  hep-ph/9405430}}]. \relax
 \relax
\bibitem{Seymour:1991xa}
M.~H. Seymour, \emph{{Photon radiation in final state parton showering}}, Z.
  Phys. \textbf{C56} (1992),
  \href{http://www-spires.dur.ac.uk/spires/find/hep/www?j=ZEPYA,C56,161}{161--%
170}. \relax
 \relax
\bibitem{Seymour:1994bx}
M.~H. Seymour, \emph{{Soft isolated photon production as a probe of the parton
  shower mechanism}}, Z. Phys. \textbf{C64} (1994),
  \href{http://www-spires.dur.ac.uk/spires/find/hep/www?j=ZEPYA,C64,445}{445--%
452}. \relax
 \relax
\bibitem{Sjostrand:2006za}
T.~Sj{\"o}strand, S.~Mrenna and P.~Skands, \emph{{PYTHIA 6.4 physics and
  manual}}, JHEP \textbf{05} (2006),
  \href{http://www.slac.stanford.edu/spires/find/hep/www?eprint=hep-ph/0603175%
}{026},  [\href{http://arXiv.org/pdf/hep-ph/0603175}{{\tt hep-ph/0603175}}].
  \relax
 \relax
\bibitem{Sjostrand:2007gs}
T.~Sj{\"o}strand, S.~Mrenna and P.~Skands, \emph{{A brief introduction to
  PYTHIA 8.1}}, Comput. Phys. Commun. \textbf{178} (2008),
  \href{http://www.slac.stanford.edu/spires/find/hep/www?eprint=arXiv:0710.382%
0}{852--867},  [\href{http://arXiv.org/pdf/0710.3820}{{\tt arXiv:0710.3820}}
  [hep-ph]]. \relax
 \relax
\bibitem{Corcella:2000bw}
G.~Corcella et~al., \emph{{HERWIG 6: an event generator for hadron emission
  reactions with interfering gluons (including supersymmetric processes)}},
  JHEP \textbf{01} (2001),
  \href{http://www-spires.dur.ac.uk/spires/find/hep/www?eprint=hep-ph/0011363}%
{010},  [\href{http://arXiv.org/pdf/hep-ph/0011363}{{\tt hep-ph/0011363}}].
  \relax
 \relax
\bibitem{Bahr:2008pv}
M.~B{\"a}hr et~al., \emph{{Herwig++ Physics and Manual}}, Eur. Phys. J.
  \textbf{C58} (2008),
  \href{http://www.slac.stanford.edu/spires/find/hep/www?eprint=0803.0883}{639%
--707},  [\href{http://arXiv.org/pdf/0803.0883}{{\tt arXiv:0803.0883}}
  [hep-ph]]. \relax
 \relax
\bibitem{Lonnblad:1992tz}
L.~L{\"o}nnblad, \emph{{Ariadne version 4: A program for simulation of QCD
  cascades implementing the colour dipole model}}, Comput. Phys. Commun.
  \textbf{71} (1992),
  \href{http://www.slac.stanford.edu/spires/find/hep/www?j=CPHCB,71,15}{15--31%
}. \relax
 \relax
\bibitem{Krauss:2004bs}
F.~Krauss, A.~Sch{\"a}licke, S.~Schumann and G.~Soff, \emph{{Simulating W / Z +
  jets production at the Tevatron}}, Phys. Rev. \textbf{D70} (2004),
  \href{http://www.slac.stanford.edu/spires/find/hep/www?eprint=hep-ph/0409106%
}{114009},  [\href{http://arXiv.org/pdf/hep-ph/0409106}{{\tt hep-ph/0409106}}].
  \relax
 \relax
\bibitem{Krauss:2005nu}
F.~Krauss, A.~Sch{\"a}licke, S.~Schumann and G.~Soff, \emph{{Simulating $W$/$Z$
  + jets production at the CERN LHC}}, Phys. Rev. \textbf{D72} (2005),
  \href{http://www.slac.stanford.edu/spires/find/hep/www?eprint=hep-ph/0503280%
}{054017},  [\href{http://arXiv.org/pdf/hep-ph/0503280}{{\tt hep-ph/0503280}}].
  \relax
 \relax
\bibitem{Gleisberg:2005qq}
T.~Gleisberg, F.~Krauss, A.~Sch{\"a}licke, S.~Schumann and J.-C. Winter,
  \emph{{Studying $W^+ W^-$ production at the Fermilab Tevatron with \Sherpa}},
  Phys. Rev. \textbf{D72} (2005),
  \href{http://www.slac.stanford.edu/spires/find/hep/www?eprint=hep-ph/0504032%
}{034028},  [\href{http://arXiv.org/pdf/hep-ph/0504032}{{\tt hep-ph/0504032}}].
  \relax
 \relax
\bibitem{Alwall:2007fs}
J.~Alwall et~al., \emph{{Comparative study of various algorithms for the
  merging of parton showers and matrix elements in hadronic collisions}}, Eur.
  Phys. J. \textbf{C53} (2008),
  \href{http://www.slac.stanford.edu/spires/find/hep/www?eprint=arXiv:0706.256%
9}{473--500},  [\href{http://arXiv.org/pdf/0706.2569}{{\tt arXiv:0706.2569}}
  [hep-ph]]. \relax
 \relax
\bibitem{Mangano:2006rw}
M.~L. Mangano, M.~Moretti, F.~Piccinini and M.~Treccani, \emph{{Matching matrix
  elements and shower evolution for top-pair production in hadronic
  collisions}}, JHEP \textbf{01} (2007),
  \href{http://www.slac.stanford.edu/spires/find/hep/www?eprint=hep-ph/0611129%
}{013},  [\href{http://arXiv.org/pdf/hep-ph/0611129}{{\tt hep-ph/0611129}}].
  \relax
 \relax
\bibitem{Alwall:2008qv}
J.~Alwall, S.~de~Visscher and F.~Maltoni, \emph{{QCD radiation in the
  production of heavy colored particles at the LHC}}, JHEP \textbf{02} (2009),
  \href{http://www.slac.stanford.edu/spires/find/hep/www?eprint=0810.5350}{017%
},  [\href{http://arXiv.org/pdf/0810.5350}{{\tt arXiv:0810.5350}} [hep-ph]].
  \relax
 \relax
\bibitem{Plehn:2008ae}
T.~Plehn and T.~M.~P. Tait, \emph{{Seeking Sgluons}}, J. Phys. \textbf{G36}
  (2009),
  \href{http://www-spires.dur.ac.uk/spires/find/hep/www?eprint=arXiv:0810.3919%
}{075001},  [\href{http://arXiv.org/pdf/0810.3919}{{\tt arXiv:0810.3919}}
  [hep-ph]]. \relax
 \relax
\bibitem{Hoeche:2009rj}
S.~H{\"o}che, F.~Krauss, S.~Schumann and F.~Siegert, \emph{{QCD matrix elements
  and truncated showers}}, JHEP \textbf{05} (2009),
  \href{http://www.slac.stanford.edu/spires/find/hep/www?eprint=arXiv:0903.121%
9}{053},  [\href{http://arXiv.org/pdf/0903.1219}{{\tt arXiv:0903.1219}}
  [hep-ph]]. \relax
 \relax
\bibitem{Schumann:2007mg}
S.~Schumann and F.~Krauss, \emph{{A parton shower algorithm based on
  Catani-Seymour dipole factorisation}}, JHEP \textbf{03} (2008),
  \href{http://www.slac.stanford.edu/spires/find/hep/www?eprint=arXiv:0709.102%
7}{038},  [\href{http://arXiv.org/pdf/0709.1027}{{\tt arXiv:0709.1027}}
  [hep-ph]]. \relax
 \relax
\bibitem{Carli:2009cg}
T.~Carli, T.~Gehrmann and S.~H{\"o}che, \emph{{Hadronic final states in
  deep-inelastic scattering with \Sherpa}}, Eur. Phys. J. \textbf{C} (2010),
  \href{http://www.slac.stanford.edu/spires/find/hep/www?rawcmd=f+eprint+0912.%
3715}{},  [\href{http://arXiv.org/pdf/0912.3715}{{\tt arXiv:0912.3715}}
  [hep-ph]]. \relax
 \relax
\bibitem{Catani:1996vz}
S.~Catani and M.~H. Seymour, \emph{{A general algorithm for calculating jet
  cross sections in NLO QCD}}, Nucl. Phys. \textbf{B485} (1997),
  \href{http://www.slac.stanford.edu/spires/find/hep/www?eprint=hep-ph/9605323%
}{291--419},  [\href{http://arXiv.org/pdf/hep-ph/9605323}{{\tt
  hep-ph/9605323}}]. \relax
 \relax
\bibitem{Catani:2002hc}
S.~Catani, S.~Dittmaier, M.~H. Seymour and Z.~Trocsanyi, \emph{{The dipole
  formalism for next-to-leading order QCD calculations with massive partons}},
  Nucl. Phys. \textbf{B627} (2002),
  \href{http://www.slac.stanford.edu/spires/find/hep/www?eprint=hep-ph/0201036%
}{189--265},  [\href{http://arXiv.org/pdf/hep-ph/0201036}{{\tt
  hep-ph/0201036}}]. \relax
 \relax
\bibitem{Platzer:2009jq}
\href{http://www-spires.dur.ac.uk/spires/find/hep/www?eprint=arXiv:0909.5593}{%
S.~Pl{\"a}tzer and S.~Gieseke}, \emph{{Coherent Parton Showers with Local
  Recoils}},  \href{http://arXiv.org/pdf/0909.5593}{{\tt arXiv:0909.5593}}
  [hep-ph]. \relax
 \relax
\bibitem{Martin:2004dh}
A.~D. Martin, R.~G. Roberts, W.~J. Stirling and R.~S. Thorne, \emph{{Parton
  distributions incorporating QED contributions}}, Eur. Phys. J. \textbf{C39}
  (2005),
  \href{http://www.slac.stanford.edu/spires/find/hep/www?eprint=hep-ph/0411040%
}{155--161},  [\href{http://arXiv.org/pdf/hep-ph/0411040}{{\tt
  hep-ph/0411040}}]. \relax
 \relax
\bibitem{Catani:1992zp}
S.~Catani, Y.~L. Dokshitzer and B.~R. Webber, \emph{{The $k_\perp$ clustering
  algorithm for jets in deep inelastic scattering and hadron collisions}},
  Phys. Lett. \textbf{B285} (1992),
  \href{http://www.slac.stanford.edu/spires/find/hep/www?j=PHLTA,B285,291}{291%
--299}. \relax
 \relax
\bibitem{Catani:1993hr}
S.~Catani, Y.~L. Dokshitzer, M.~H. Seymour and B.~R. Webber,
  \emph{{Longitudinally-invariant $k_\perp$-clustering algorithms for
  hadron--hadron collisions}}, Nucl. Phys. \textbf{B406} (1993),
  \href{http://www.slac.stanford.edu/spires/find/hep/www?j=NUPHA,B406,187}{187%
--224}. \relax
 \relax
\bibitem{Gleisberg:2003xi}
T.~Gleisberg, S.~H{\"o}che, F.~Krauss, A.~Sch{\"a}licke, S.~Schumann and
  J.~Winter, \emph{{\Sherpa 1.$\alpha$, a proof-of-concept version}}, JHEP
  \textbf{02} (2004),
  \href{http://www.slac.stanford.edu/spires/find/hep/www?irn=5730570}{056},
  [\href{http://arXiv.org/pdf/hep-ph/0311263}{{\tt hep-ph/0311263}}]. \relax
 \relax
\bibitem{Gleisberg:2008ta}
T.~Gleisberg, S.~H{\"o}che, F.~Krauss, M.~Sch\"{o}nherr, S.~Schumann,
  F.~Siegert and J.~Winter, \emph{{Event generation with \Sherpa 1.1}}, JHEP
  \textbf{02} (2009),
  \href{http://www.slac.stanford.edu/spires/find/hep/www?eprint=0811.4622}{007%
},  [\href{http://arXiv.org/pdf/0811.4622}{{\tt arXiv:0811.4622}} [hep-ph]].
  \relax
 \relax
\bibitem{Buskulic:1995au}
D.~Buskulic et~al., ALEPH collaboration, \emph{{First measurement of the
  quark-to-photon fragmentation function}}, Z. Phys. \textbf{C69} (1996),
  \href{http://www-spires.slac.stanford.edu/spires/find/hep/www?j=ZEPYA,C69,36%
5}{365--378}. \relax
 \relax
\bibitem{Aaltonen:2009ty}
T.~Aaltonen, CDF collaboration, \emph{{Measurement of the inclusive isolated
  prompt photon cross section in $p\bar{p}$ collisions at $\sqrt{s} = 1.96$ TeV
  using the CDF detector}}, Phys. Rev. \textbf{D80} (2009),
  \href{http://www-spires.dur.ac.uk/spires/find/hep/www?eprint=arXiv:0910.3623%
}{111106},  [\href{http://arXiv.org/pdf/0910.3623}{{\tt arXiv:0910.3623}}
  [hep-ex]]. \relax
 \relax
\bibitem{Abazov:2005wc}
V.~M. Abazov et~al., D0 collaboration, \emph{{Measurement of the isolated
  photon cross section in $p \bar{p}$ collisions at $\sqrt{s}$ = 1.96 TeV}},
  Phys. Lett. \textbf{B639} (2006),
  \href{http://www.slac.stanford.edu/spires/find/hep/www?eprint=hep-ex/0511054%
}{151--158},  [\href{http://arXiv.org/pdf/hep-ex/0511054}{{\tt
  hep-ex/0511054}}]. \relax
 \relax
\bibitem{Acosta:2004sn}
D.~E. Acosta et~al., CDF collaboration, \emph{{Measurement of the Cross Section
  for Prompt Diphoton Production in $p\bar{p}$ Collisions at $\sqrt{s} = 1.96$
  TeV}}, Phys. Rev. Lett. \textbf{95} (2005),
  \href{http://www-spires.dur.ac.uk/spires/find/hep/www?eprint=hep-ex/0412050}%
{022003},  [\href{http://arXiv.org/pdf/hep-ex/0412050}{{\tt hep-ex/0412050}}].
  \relax
 \relax
\bibitem{Chen:1997gg}
W.~Chen, \emph{{Isolated direct double photon production in $p\bar{p}$
  collisions at 1.8-TeV with the D0 detector}}, FERMILAB-THESIS-1997-31. \relax
 \relax
\bibitem{Aivazis:1993pi}
M.~A.~G. Aivazis, J.~C. Collins, F.~I. Olness and W.-K. Tung,
  \emph{{Leptoproduction of heavy quarks. II. A unified QCD formulation of
  charged and neutral current processes from fixed-target to collider
  energies}}, Phys. Rev. \textbf{D50} (1994),
  \href{http://www-spires.dur.ac.uk/spires/find/hep/www?eprint=hep-ph/9312319}%
{3102--3118},  [\href{http://arXiv.org/pdf/hep-ph/9312319}{{\tt
  hep-ph/9312319}}]. \relax
 \relax
\bibitem{Pittau:1996ez}
R.~Pittau, \emph{{A simple method for multi-leg loop calculations}}, Comput.
  Phys. Commun. \textbf{104} (1997),
  \href{http://www.slac.stanford.edu/spires/find/hep/www?eprint=hep-ph/9607309%
}{23--36},  [\href{http://arXiv.org/pdf/hep-ph/9607309}{{\tt hep-ph/9607309}}].
  \relax
 \relax
\bibitem{Pittau:1997mv}
R.~Pittau, \emph{{A simple method for multi-leg loop calculations 2: a general
  algorithm}}, Comput. Phys. Commun. \textbf{111} (1998),
  \href{http://www.slac.stanford.edu/spires/find/hep/www?eprint=hep-ph/9712418%
}{48--52},  [\href{http://arXiv.org/pdf/hep-ph/9712418}{{\tt hep-ph/9712418}}].
  \relax
 \relax
\bibitem{Gleisberg:2008fv}
T.~Gleisberg and S.~H{\"o}che, \emph{{Comix, a new matrix element generator}},
  JHEP \textbf{12} (2008),
  \href{http://www.slac.stanford.edu/spires/find/hep/www?rawcmd=f+eprint+0808.%
3674}{039},  [\href{http://arXiv.org/pdf/0808.3674}{{\tt arXiv:0808.3674}}
  [hep-ph]]. \relax
 \relax
\bibitem{Pumplin:2002vw}
J.~Pumplin, D.~R. Stump, J.~Huston, H.~L. Lai, P.~Nadolsky and W.~K. Tung,
  \emph{{New generation of parton distributions with uncertainties from global
  QCD analysis}}, JHEP \textbf{0207} (2002),
  \href{http://www.slac.stanford.edu/spires/find/hep/www?eprint=hep-ph/0201195%
}{012},  [\href{http://arXiv.org/pdf/hep-ph/0201195}{{\tt hep-ph/0201195}}].
  \relax
 \relax
\bibitem{Winter:2003tt}
J.-C. Winter, F.~Krauss and G.~Soff, \emph{{A modified cluster-hadronisation
  model}}, Eur. Phys. J. \textbf{C36} (2004),
  \href{http://www.slac.stanford.edu/spires/find/hep/www?eprint=hep-ph/0311085%
}{381--395},  [\href{http://arXiv.org/pdf/hep-ph/0311085}{{\tt
  hep-ph/0311085}}]. \relax
 \relax
\bibitem{Krauss:2010xy}
\emph{{\Sherpa's new hadronisation model}}, in preparation. \relax
 \relax
\bibitem{Krauss:2010xx}
F.~Krauss, T.~Laubrich and F.~Siegert, \emph{{Simulation of hadron decays in
  \Sherpa}}, in preparation. \relax
 \relax
\bibitem{Buckley:2009bj}
A.~Buckley, H.~Hoeth, H.~Lacker, H.~Schulz and J.~E. von Seggern,
  \emph{{Systematic event generator tuning for the LHC}}, Eur. Phys. J.
  \textbf{C65} (2010),
  \href{http://www.slac.stanford.edu/spires/find/hep/www?eprint=arXiv:0907.297%
3}{331--357},  [\href{http://arXiv.org/pdf/0907.2973}{{\tt arXiv:0907.2973}}
  [hep-ph]]. \relax
 \relax
\bibitem{Schonherr:2008av}
M.~Sch\"{o}nherr and F.~Krauss, \emph{Soft photon radiation in particle decays
  in \Sherpa}, JHEP \textbf{12} (2008),
  \href{http://www.slac.stanford.edu/spires/find/hep/www?eprint=arXiv:0810.507%
1}{018},  [\href{http://arXiv.org/pdf/0810.5071}{{\tt arXiv:0810.5071}}
  [hep-ph]]. \relax
 \relax
\end{thebibliography}
\end{document}